%% file: main.tex
\documentclass[conference]{IEEEtran}
\IEEEoverridecommandlockouts

\usepackage{cite}
\usepackage{amsmath,amssymb,amsfonts}
\usepackage{algorithmic}
\usepackage{graphicx}
\usepackage{textcomp}
\usepackage{xcolor}
\def\BibTeX{{\rm B\kern-.05em{\sc i\kern-.025em b}\kern-.08em
    T\kern-.1667em\lower.7ex\hbox{E}\kern-.125emX}}

\usepackage{tikz}
\usepackage[colorinlistoftodos]{todonotes}
\usepackage{booktabs,pifont,array,tabularx}
\newcommand{\cmark}{\ding{51}}  \newcommand{\xmark}{\ding{55}}

\usepackage{url}
\usepackage{hyperref}
\usepackage{makecell} 
\usepackage{multirow} 
\usepackage{siunitx}
\usepackage{listings}
\usepackage{subcaption}
\usepackage{tcolorbox}
\usepackage{enumitem}

\usepackage{rotating}   
\usepackage{graphicx}   

\begin{document}

\title{SoK: Are LLMs Reliable at Source Code Recovery? A Taxonomy and Empirical Evaluation}


\author{
\IEEEauthorblockN{
Varun Kohli\IEEEauthorrefmark{1},
Lee Bing Cheng\IEEEauthorrefmark{1},
Nur Hazim Ghazali\IEEEauthorrefmark{2},
Gao Yuze\IEEEauthorrefmark{1},
Daryl Poon\IEEEauthorrefmark{1},
and Dinil Mon Divakaran\IEEEauthorrefmark{1}}
\IEEEauthorblockA{\IEEEauthorrefmark{1}A*STAR Institute of Advanced Intelligence and Computing, Singapore\\
\{Kohli\_Varun, Lee\_Bing\_Cheng, Gao\_Yuze, Daryl\_Poon, Dinil\_Divakaran\}@a-star.edu.sg}
\IEEEauthorblockA{\IEEEauthorrefmark{2}National University of Singapore, Singapore\\
nurhazim@u.nus.edu}
}

\maketitle

\begin{abstract}
Effective source recovery is critical to security applications such as malware analysis, vulnerability assessment, and legacy maintenance. Large Language Models (LLMs) are reshaping this field, shifting the paradigm away from rule-based heuristics to probabilistic and high fidelity semantic recovery of source code from assembly or classical decompiler-derived pseudo-C. However, despite rapid progress, the field suffers from  fragmentation across numerous approaches as well as their non-unified evaluations, limiting objective comparisons. Further, existing works have limited coverage of embedded, IoT architectures and source languages beyond C/C++. In this work, we present the first Systematization of Knowledge (SoK) focused specifically on LLM-assisted binary-to-source recovery. We provide a granular design-centric taxonomy of LLM-assisted source recovery methods and systematic evaluations using seven key metrics along six evaluation dimensions. We evaluate state-of-the-art methods on 45,000 test samples derived from four standard and five embedded architectures, five optimization levels, and symbol stripping. We ablate the impact of design choices on recovery performance, including input representation, contextual enrichment, model scale, iteration and review roles using controlled in-house recovery pipelines and three off-the-shelf models. Finally, we test same-language and cross-language recovery capability covering four mature and two legacy languages. Our systematization and comprehensive evaluations provide key insights that guide future directions in this field.
\end{abstract}

\begin{IEEEkeywords}
Source Code Recovery, Large Language Models, Reliability, Reverse Engineering, Binary Decompilation
\end{IEEEkeywords}

\section{Introduction}
Source code recovery is a critical building block of reverse engineering, computer security, and software maintenance. It enables security analysts to inspect and investigate programs~\cite{investigate-process-2020}, analyze malware~\cite{psybot-2013}, audit third-party binaries for vulnerabilities~\cite{shoshitaishvili2015firmalice}, plagiarism detection~\cite{graph-BCS-2017}, and so on, all without access to original source code. Traditional decompilers such as Hex-Rays~\cite{hexrays_decompiler}, Binary Ninja~\cite{binary-ninja}, Ghidra~\cite{ghidra}, and Angr~\cite{angr} typically operate on a pipeline of deterministic Intermediate Representations (IR), control flow reconstruction algorithms, pattern matching and type inference heuristics. While these classical tools preserve lower-level control execution semantics, their output is marred by issues like poor readability, unnatural control constructs, lost variable names, and fragmented type structures, which ultimately hinders human analysis and recompilation~\cite{taxonomy-fidelity-2024}.

The emergence of Large Language Models (LLMs) trained on massive language and code corpora has led to unprecedented capabilities in code generation, analysis, and software comprehension. Building upon this proficiency, LLMs are now increasingly being used to address the challenges in reverse engineering \cite{WaDec, ReSym-CCS-2024, jiang2025beyond}.
Rather than relying purely on deterministic rules, LLM-assisted source recovery models treat decompilation or pseudo-C refinement as a neural sequence-to-sequence translation task that converts low-level representations to recompilable and equivalent high-level source code. These approaches show promise in recovering idiomatic function names, expressive variable and type structures, and clean high-level abstractions directly from assembly or decompiled pseudo-C output~\cite{ReSym-CCS-2024, xu2025unleashing, tan2024llm4decompile, func-name-pred-TSEM, typeforge-SP2025, jiang2025nova, tian2026decompile}. 
Despite rapid proliferation across top security, software engineering, and AI venues, the research landscape remains highly fragmented. Even when restricting the scope to top-tier venues (see Section~\ref{subsec:literature-scan-method}) over the past two years, 13 distinct methodologies have been published~\cite{hu2024degpt,tan2024llm4decompile,feng2024self,wong2025decllm,sun2025enhancing,jiang2025nova,li2025pseudofix,cao2026behavior,zhou2025fidelitygpt,dramko2026idioms,abualazm2026llms,li2026sk2decompile,su2026rldecompiler}, alongside several pre-prints this year~\cite{zhang2026constraint,cui2026pcodetrans,liu2026binary,shypula2026decaf,gagnon2026practical,kohli2026chisel}. 

Prior systematizations broadly explore LLM applications in reverse engineering with coarse taxonomies of LLM-based source recovery~\cite{hu2025sok, abusabha2026evolution, liu2026llm}. To the best of our knowledge, there is no existing systematization of the granular design choices of LLM-assisted source recovery methods, and no controlled ablations of such choices to measure their impact on fidelity. Further, evaluation metrics are highly fragmented, with $53$ unique metrics identified across 20+ research papers in top venues on LLM-assisted source recovery from 2024--2026. Moreover, existing evaluations and benchmarks on LLM-assisted source recovery focus almost exclusively on x86-64 and C/C++ target generation \cite{gao2025decompilebench,tian2026decompile,yoda2026statisticalanalysisexecutabilityprogram}, with coverage of non-C languages only by a few method papers and limited to C++ \cite{wong2025decllm, li2025pseudofix}, CUDA \cite{sun2025enhancing}, and Dart \cite{abualazm2026evaluating}. Further, decompilation fidelity across legacy and mature languages has only been studied for pre-LLM neural decompiations \cite{Hosseini2022BeyondTC}. There is no existing evaluation of LLM-assisted source recovery that covers embedded systems architectures (armv7m, armv8m, armv7a, armv8a, and mips23le) \cite{tan2024sok}.  

To bridge this gap, this paper presents the first comprehensive Systematization of Knowledge (SoK) dedicated to LLM-assisted source recovery. We formulate four core Research Questions (RQs) to evaluate state-of-the-art (SOTA) decompilation paradigms:

\begin{itemize}
    \item \textbf{RQ1: Evaluation Dimensions} (Sections \ref{sec:metrics}, \ref{subsec:metrics}, \ref{subsec:sota_eval}). As the \textit{Law of Leaky Abstraction} (Joel Spolsky) states, ``All non-trivial abstractions, to some extent, are leaky." Two distinct source codes may have the same functional behavior through the reduction of noise to machine language during compilation. Thus, what dimensions must be considered to evaluate the fidelity of source recovery?
    \item \textbf{RQ2: Generalization} (Section \ref{subsec:sota_eval}). How well does SOTA LLM-assisted source recovery generalize when evaluated against cross-architecture binaries covering standard and embedded (IoT) architectures, diverse compiler optimization flags (\texttt{O0}-\texttt{Os}), and symbol availability?
    \item \textbf{RQ3: Pipeline Design} (Sections \ref{subsec:literature-scan-method}, \ref{subsec:sota_eval}, \ref{subsec:taxonomy_ablation}). What are the key building blocks and design choices of LLM-assisted methods, and their impact on source recovery?
    \item \textbf{RQ4: Multi-Language Recovery Spectrum} (Section \ref{subsec:language_study}). How effectively can LLMs recover the same language or cross-recover C from a different mature or legacy source language?
\end{itemize}

The following are the key findings from our evaluations to answer the above questions (see Section~\ref{sec:key-findings-OP} for details):

\begin{tcolorbox}[
    colback=orange!5,
    colframe=orange!60!brown,
    title=Key Findings of Our SoK,
    fonttitle=\bfseries,
]
\begin{itemize}[leftmargin=*]
    \item \textbf{Best SOTA:} As of this study, Agent4Decompile \cite{zhang2026constraint} is the best method for recovering compilabile code with byte-level semantics and functional similarity. SK2Decompile \cite{li2026sk2decompile} and FidelityGPT \cite{zhou2025fidelitygpt} follow behind, outperforming mainly in relative readability relative to pseudo-C and lexical similarity. 
    \item \textbf{Signature recovery:} Function-level recompilation rates hide signature recovery problems, making recompilability with call interfaces a better metric for reporting as well as iterative correction. 
    \item \textbf{Generalization:} SOTA methods perform worse on Thumb2 (IoT) ISAs than standard ISAs. Performance drops on aggressive optimization, while symbol stripping mainly impacts readability.
    \item \textbf{Design choices:} Pseudo-C acts as a portability layer for cross-architecture recovery due to standardized output of classical decompilers. Compiler iterations and larger models improve compilability, binary-level semantics and functional recovery, while LLM roles improve readability.
    \item \textbf{Non-C recovery:} Current LLM and classical decompiler capabilities are C-centric and heavily under-perform in non-C recovery.
\end{itemize}
\end{tcolorbox}

\textbf{Contributions:} In summary, we make the following contributions through this work:

\begin{enumerate}
    \item A systematization of LLM-assisted source recovery methods in top cybersecurity, AI, and software engineering venues, and a new taxonomy derived across five orthogonal axes, namely, model setup, control-flow, number of model roles per turn, input representations, and augmentation. The design-centric taxonomy enables a systematic evaluation of SOTA and targeted ablations.
    \item A systematization of evaluation metrics used in prior studies highlights 53 metrics used across recent research on LLM-assisted source recovery across six key dimensions: two quality (compilation, readability) and four similarity (lexical, syntactic, semantic, functional). 7 metrics are selected across the 6 axes to enable a holistic, multi-axes evaluation across SOTA and ablations.
    \item A comprehensive evaluation of SOTA methods on 45,000 samples derived from \textit{four standard} and \textit{five embedded (IoT)} architectures, five optimization levels, and unstripped or stripped symbol variants. 
    \item A controlled, taxonomy-driven ablation using three off-the-shelf Gemma4 (12b-it, 31b-it) \cite{gemmateam2026gemma4} and Qwen3.8-27b \cite{qwen38} models, various base input representations and enrichment channels, and in-house source recovery pipelines including single-pass and single-roles (SP), compiler iteration (CIT), and review role (RR) to evaluate the impact of various base input representations, contextual enrichment, model scale, feedback iterations, and review turns on source recovery. 
    \item An evaluation of two off-the-shelf LLMs in same-language recovery and cross-language recovery (to C) from binary tasks across mature (C, C++, Go, Rust) and legacy (Fortran, Ada) languages.  
\end{enumerate}

\section{Preliminaries}
\label{sec:preliminaries}
To contextualize the shift from deterministic heuristics to neural source code recovery, this section outlines the fundamental mechanics of information loss during compilation and the differences between classical and neural decompilers.

\subsection{The Lossy Nature of Compilation}
Compilation is an inherently destructive, one-way transformation where a decompiler cannot recover the original source deterministically \cite{dramko2026codealign}. When high-level source code is translated into a machine-executable binary, the compiler strips away rich semantic abstractions such as meaningful variable names, custom data structures, class hierarchies, and idiomatic control flow constructs \cite{learned_variable_names_types, class_hierarchies, recover_data_structures, lee2011tie}. 
Furthermore, aggressive compiler optimizations can significantly refactor the underlying program logic to maximize hardware efficiency, through, for example, loop unrolling, function inlining, and dead code elimination \cite{bin2wrong, loop_unrolling, abusabha2025deep_function_inlining, aho2006compilers}. Consequently, recovering the original source code is a challenging task, rooted in the many-to-one nature of the compilation process~\cite{binary-not-easy-2016, disassembly-undecidability-2024, Trex-USENIX-SEC-2025}. The practical goal of decompilation is therefore to recover a functionally equivalent and human-readable representation for a variety of use cases that include automated and manual analysis for vulnerability detection, patch diffing, malware analysis, program comprehension, and software license compliance \cite{decompile_optimized_binary, tan2024llm4decompile, decompiler_effectiveness}.

\subsection{Classical Decompilation Mechanics}
Traditional decompilers (e.g., Hex-Rays, Ghidra) address binary recovery through deterministic, iterative pipelines \cite{d-helix}. They typically disassemble raw machine code, lift the assembly instructions into an Intermediate Representation (IR), and perform extensive data-flow and control-flow graph (CFG) analyses \cite{ghidra_sleigh, Yakdan2015NoMG, phoenix_decompiler}. Finally, heuristic-based structural analysis and type inference are applied to emit C-like pseudo-C. While mathematically sound and highly effective at preserving low-level execution semantics, these tools are brittle. Failure could occur when encountering heavily optimized code, non-standard calling conventions, or obfuscated binaries, often producing pseudo-C cluttered with unnatural \texttt{goto} statements and fragmented types that inhibit human comprehension \cite{decompiler_effectiveness, SAILR, bin2wrong}.

\subsection{The Neural Decompiler Paradigm}
The integration of Large Language Models (LLMs) frames source code recovery not as a problem of deterministic graph reduction, but as a neural translation task. By leveraging transformer architectures pre-trained on vast corpora of language, code, and assembly, neural decompilers attempt to probabilistically infer the lost semantic context. Instead of relying on hardcoded heuristics, these models ingest assembly language \cite{jiang2025nova, tan2024llm4decompile, feng2024self}, or pseudo-C \cite{tan2024llm4decompile} emitted by classical decompilers, and output high-level source code. Because they evaluate the global context of the input sequence, LLMs excel at generating idiomatic function names, reconstructing complex nested logic, and emitting highly readable code \cite{WaDec}. However, this probabilistic nature introduces a critical vulnerability: hallucinations. Neural models may generate syntactically flawless code that subtly alters the functional behavior of the original binary, creating a tension between readability and semantic or functional determinism \cite{gao2025decompilebench}.


\begin{figure}
    \centering
    \includegraphics[width=0.8\linewidth]{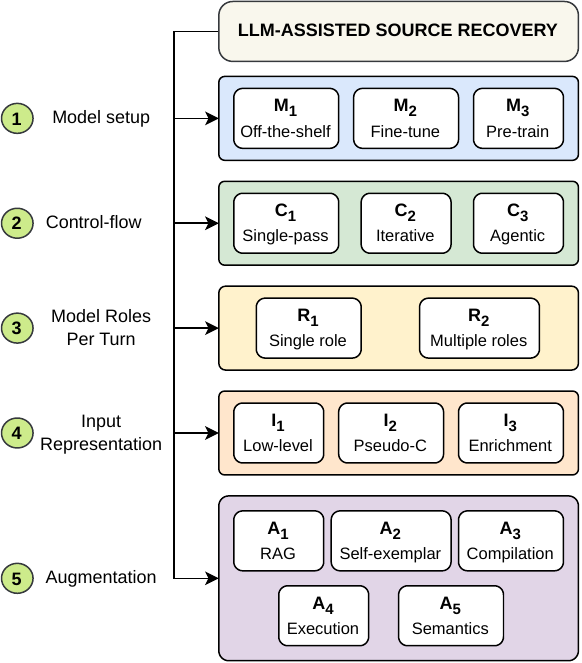}
    \caption{Taxonomy of LLM-assisted methods along five axes - model setup (M), control flow (C), model roles per turn (R), input representations (I), and augmentation (A).}
    \label{fig:taxonomy}
\end{figure}

\section{Systematization}
\label{sec:systematization}

This section dives into the current literature on LLM-assisted source recovery and derive a five-axes taxonomy for this field. We gathered our literature corpus of research papers by querying DBLP over Jan 2020 to July 2026 with the keyword set (and variants) consisting of \emph{``decompilation"}, 
\emph{``binary to source"}, \emph{``assembly to source"}, \emph{bytecode to source}, \emph{``pseudocode to source"}, \emph{``source code recovery"}, and \emph{``recover* source code"}. This search was restricted to A and A* CORE ranking\footnote{https://portal.core.edu.au/conf-ranks/} venues across three fields: \textbf{\emph{Cybersecurity}} (S\&P, USENIX Security, CCS, NDSS, EuroS\&P, ASIACCS, CSF, ESORICS, RAID, ACSAC, DSN, TIFS, TDSC, WWW), \textbf{\emph{Software Engineering}} (ICSE, FSE/ISSTA, ASE, SANER, ICSME, ISSRE, MSR, ICPC, ESEM, TSE, TOSEM, EMSE), and \textbf{\emph{AI}} (NeurIPS, ICML, ICLR, AAAI, IJCAI, ACL, EMNLP, NAACL, COLING, EACL, COLM, AISTATS, UAI, TACL, JMLR, TMLR). We manually added two missing works, i.e., Nova~\cite{jiang2025nova}, due to the lack of relevant keywords in its title, and SK2Decompile~\cite{li2026sk2decompile}, due to recency of publication. This approach returned $57$ candidate papers, of which $52$ are relevant to decompilation consisting of $35$ methods and $17$ surveys or evaluations. Among the $35$ method papers, $31$ lie specifically in the scope of source-recovery, breaking down further into $19$ LLM-assisted and $12$ classical (non-LLM) works on source recovery. We focus only on the LLM-assisted methods, excluding six which target Virtual-Machine (VM) bytecode such as WebAssembly~\cite{WaDec,3692070.3692591} and blockchain smart contracts \cite{su2025disco,giesen2025poster,chen2025suigpt,liao2025augmenting}, since that lies outside our standard and embedded computer-architecture scope. The resultant $13$ peer-reviewed in-scope methods are augmented by $5$ arXiv pre-prints until July 2026 selected using the same methodology to create an exhaustive and diverse set of $18$ LLM-assisted source-recovery methods included chronologically in Table~\ref{tab:sota}. Following the same funneling logic as above, we consider four benchmarks and empirical evaluations of LLM-assisted recovery to source code; we cover these in the next section.\footnote{While there are LLM-based research  that focus on specific aspects of decompilation, e.g., variable names, types and function name prediction~\cite{ReSym-CCS-2024, xu2025unleashing, typeforge-SP2025, func-name-pred-TSEM}, and works on compilers, IR and pseudo-C \cite{zhu-etal-2026-lifting,armengol2024forklift,cummins2025llm}, we limit the scope of our work to the specific problem of ``source code recovery". The following subsections compile the surveyed benchmark and evaluation studies, and then systematize method papers on LLM-assisted source recovery.} 

\begin{table*}[t]
\centering
\caption{LLM-assisted decompilation (source-code recovery) methods for standard/embedded computer architectures in top (A/A*) cybersecurity, AI, and software-engineering venues, 2020--2026, followed by 2026 arXiv preprints.}
\label{tab:sota}
\scriptsize
\renewcommand{\arraystretch}{1.25}
\begin{tabular*}{\textwidth}{@{\extracolsep{\fill}}|l|l|c|c|c|c|c|>{\raggedright\arraybackslash}p{1.4cm}|>{\raggedright\arraybackslash}p{0.9cm}|>{\raggedright\arraybackslash}p{1.2cm}|>{\raggedright\arraybackslash}p{0.7cm}|c|}
\hline
\textbf{Reference} & \textbf{Venue} \& \textbf{Year} & \textbf{Model} & \textbf{Control} & \textbf{Role} & \textbf{Input} & \textbf{Augment} & \textbf{Architecture} & \textbf{Opt} & \textbf{Symbols} & \textbf{Source} & \textbf{Code} \\ \hline
DeGPT~\cite{hu2024degpt} & NDSS 2024 & $\text{M}_1$ & $\text{C}_1$ & $\text{R}_2$ & $\text{I}_2$ & $\text{A}_5$ & Not reported & \texttt{O0-O3} & stripped, unstripped & C & \cmark \\ \hline
LLM4Decompile~\cite{tan2024llm4decompile} & EMNLP 2024 & $\text{M}_2$ & $\text{C}_1$ & $\text{R}_1$ & $\text{I}_1$, $\text{I}_2$ & --- & \texttt{x86-64} & \texttt{O0-O3} & unstripped & C & \cmark \\ \hline
SCCDEC~\cite{feng2024self} & EMNLP 2024 & $\text{M}_1$, $\text{M}_2$ & $\text{C}_1$ & $\text{R}_1$ & $\text{I}_1$ & $\text{A}_2$ & \texttt{x86-64} & \texttt{O0-O3} & Not reported & C & \cmark \\ \hline
DecLLM~\cite{wong2025decllm} & ISSTA 2025 & $\text{M}_1$ & $\text{C}_2$ & $\text{R}_1$ & $\text{I}_2$ & $\text{A}_3$, $\text{A}_4$ & Not reported & \texttt{O0-O3} & stripped, unstripped & C, C++ & \xmark \\ \hline
PtxDec~\cite{sun2025enhancing} & ASE 2025 & $\text{M}_2$ & $\text{C}_1$ & $\text{R}_1$ & $\text{I}_1$ & --- & PTX & TVM scheduler & Not reported & CUDA & \cmark \\ \hline
Nova~\cite{jiang2025nova} & ICLR 2025 & $\text{M}_3$ & $\text{C}_1$ & $\text{R}_1$ & $\text{I}_1$ & --- & \texttt{x86-64} & \texttt{O0-O3} & stripped & C & \cmark \\ \hline
PseudoFix~\cite{li2025pseudofix} & ASE 2025 & $\text{M}_1$ & $\text{C}_1$ & $\text{R}_2$ & $\text{I}_2$ & $\text{A}_1$ & \texttt{x86-64} & \texttt{O0-O3} & stripped & C/C++ & \cmark \\ \hline
Cao ~\cite{cao2026behavior} & DSN 2026 & $\text{M}_1$ & $\text{C}_3$ & $\text{R}_1$ & $\text{I}_1$, $\text{I}_3$ & $\text{A}_3$, $\text{A}_4$ & \texttt{AVR} & \texttt{Os} & stripped & C & \xmark \\ \hline
FidelityGPT~\cite{zhou2025fidelitygpt} & NDSS 2026 & $\text{M}_1$ & $\text{C}_1$ & $\text{R}_2$ & $\text{I}_2$, $\text{I}_3$ & $\text{A}_1$ & \texttt{x86 (32/64), ARM-32, MIPS-32} & \texttt{O0-O3} & Not reported & C & \cmark \\ \hline
Idioms~\cite{dramko2026idioms} & NDSS 2026 & $\text{M}_2$ & $\text{C}_1$ & $\text{R}_1$ & $\text{I}_2$, $\text{I}_3$ & --- & \texttt{x86-64} & \texttt{O0-O3} & stripped & C & \cmark \\ \hline
Abualazm et al. \cite{abualazm2026llms} & SANER 2026 & $\text{M}_2$ & $\text{C}_1$ & $\text{R}_1$ & $\text{I}_1$ & --- & \texttt{x86-64} & AOT & Not reported & Dart & \cmark \\ \hline
SK2Decompile~\cite{li2026sk2decompile} & ICLR 2026 & $\text{M}_2$ & $\text{C}_1$ & $\text{R}_2$ & $\text{I}_2$ & --- & \texttt{x86-64} & \texttt{O0-O3} & stripped & C & \cmark \\ \hline
RlDecompiler~\cite{su2026rldecompiler} & ICPC 2026 & $\text{M}_2$ & $\text{C}_1$ & $\text{R}_1$ & $\text{I}_1$, $\text{I}_3$ & --- & \texttt{x86-64} & \texttt{O0-O3}, \texttt{Os} & not reported & C & \cmark \\ \hline
Agent4Decompile~\cite{zhang2026constraint} & arXiv 2026 & $\text{M}_1$ & $\text{C}_2$ & $\text{R}_2$ & $\text{I}_2$ & $\text{A}_3$, $\text{A}_4$ & \texttt{x86-64} & \texttt{O0-O3} & stripped, unstripped & C & \cmark \\ \hline
PCodeTrans~\cite{cui2026pcodetrans} & arXiv 2026 & $\text{M}_1$ & $\text{C}_2$ & $\text{R}_2$ & $\text{I}_2$, $\text{I}_3$ & $\text{A}_3$, $\text{A}_4$ & \texttt{x86-64} & \texttt{O2} & stripped, unstripped & C & \xmark \\ \hline
AutoDecompiler~\cite{liu2026binary} & arXiv 2026 & $\text{M}_2$ & $\text{C}_2$ & $\text{R}_1$ & $\text{I}_1$, $\text{I}_2$ & $\text{A}_3$, $\text{A}_4$ & \texttt{x86-64} & \texttt{O0-O3} & Not reported & C & \xmark \\ \hline
Decaf~\cite{shypula2026decaf} & arXiv 2026 & $\text{M}_2$ & $\text{C}_1$ & $\text{R}_1$ & $\text{I}_2$, $\text{I}_3$ & $\text{A}_3$ & \texttt{x86-64} & \texttt{O0, O2} & stripped, unstripped & C & \cmark \\ \hline
Gagnon et al.~\cite{gagnon2026practical} & arXiv 2026 & $\text{M}_1$ & $\text{C}_2$ & $\text{R}_1$ & $\text{I}_1$, $\text{I}_2$, $\text{I}_3$ & $\text{A}_1$ & \texttt{x86-64} & \texttt{O2} & stripped & C & \xmark \\ \hline
\end{tabular*}
\end{table*}

\subsection{Benchmarks and Empirical Evaluations}
We first evaluate the coverage of existing surveys and evaluations of LLM-assisted source code recovery and highlight the gap in systematization. 

Gao {et al.}~\cite{gao2025decompilebench} benchmark twelve traditional and LLM decompilers on $23{,}400$ x86-64 functions from $130$ OSS-Fuzz C projects, reporting recompilation success, a runtime coverage-equivalence rate, and a twelve-dimensional LLM-judged quality score across optimization levels \texttt{O0-O3} and \texttt{Os}. Another benchmark~\cite{tian2026decompile} contributes a million-scale corpus of two million binary-source function pairs mined from permissively licensed C and C++ repositories, evaluating re-executability, R2I readability, and edit similarity over \texttt{O0-O3}. Abualazm et al.~\cite{abualazm2026evaluating} study fine-tuning and metric validity for Dart binaries, showing on a new benchmark that CodeBLEU and compile@$k$ diverge from functional pass@$k$, and are the only work to examine cross-language transfer, from Swift to Dart. Yoda et al.~\cite{yoda2026statisticalanalysisexecutabilityprogram} propose a nine-dimensional structural, behavioral, and semantic quality metric and analyze roughly nineteen thousand decompilations of $318$ router programs compiled to x86-64 across \texttt{O0, O2, O3}, and \texttt{Os}.

Across these works, evaluations are conducted on x86-64 alone. None of them ablates the input representation: each fixes a single form, whether raw assembly or classical decompiler pseudo-C, rather than comparing the variety of input representations that existing methods depend on. Language coverage is almost entirely C and C++~\cite{gao2025decompilebench,tian2026decompile,yoda2026statisticalanalysisexecutabilityprogram}. Cross-language recovery examined only between Dart and Swift~\cite{abualazm2026evaluating} and no evaluation of other mature languages such as Go, Rust, Cython, Java, and Python, and legacy ones such as Fortran.

\subsection{Systematization of LLM-assisted Source Recovery Approaches in the Literature}

\label{subsec:literature-scan-method}
We characterize each research work through five-axes taxonomy, as depicted in Fig.~\ref{fig:taxonomy} and Table~\ref{tab:sota}: 
\begin{enumerate}
    \item \textbf{Model setup (M):} This axis depicts how the underlying LLM is prepared: off-the-shelf ($\text{M}_1$) \cite{hu2024degpt,wong2025decllm,li2025pseudofix,cao2026behavior,zhou2025fidelitygpt,zhang2026constraint,cui2026pcodetrans,gagnon2026practical}, fine-tuned from a checkpoint ($\text{M}_2$) \cite{tan2024llm4decompile,feng2024self,sun2025enhancing,dramko2026idioms,abualazm2026llms,li2026sk2decompile,liu2026binary,shypula2026decaf,su2026rldecompiler}, or pre-trained for decompilation ($\text{M}_3$) \cite{jiang2025nova}.
    \item \textbf{Control flow (C):} This axis characterizes how the model is driven, including single-pass generation ($\text{C}_1$) \cite{hu2024degpt,tan2024llm4decompile,feng2024self,sun2025enhancing,jiang2025nova,li2025pseudofix,zhou2025fidelitygpt,dramko2026idioms,abualazm2026llms,li2026sk2decompile,shypula2026decaf,su2026rldecompiler}, iterative feedback-driven repair loop ($\text{C}_2$) \cite{wong2025decllm,zhang2026constraint,cui2026pcodetrans,liu2026binary,gagnon2026practical}, or an autonomous tool-using agent ($\text{C}_3$) \cite{cao2026behavior}.
    \item \textbf{Number of roles per turn (R):} This axis depicts the functional decomposition of the LLM into various roles, for example, ``code generator" and ``reviewer", per decompilation pass. Types including single role ($\text{R}_1$) \cite{tan2024llm4decompile,feng2024self,wong2025decllm,jiang2025nova,sun2025enhancing,dramko2026idioms,abualazm2026llms,cao2026behavior,liu2026binary,shypula2026decaf,gagnon2026practical,su2026rldecompiler}, or multiple specialized roles ($\text{R}_2$) \cite{hu2024degpt,li2025pseudofix,zhou2025fidelitygpt,li2026sk2decompile,zhang2026constraint,cui2026pcodetrans}.
    \item \textbf{Input representation (I):} This axis depicts the binary-derived forms included in the input to the LLM, covering forms such has hex, assembly, and low-level IR ($\text{I}_1$) \cite{tan2024llm4decompile,feng2024self,sun2025enhancing,jiang2025nova,cao2026behavior,abualazm2026llms,liu2026binary,gagnon2026practical,su2026rldecompiler}, classical decompiler pseudo-C ($\text{I}_2$) \cite{hu2024degpt,tan2024llm4decompile,wong2025decllm,li2025pseudofix,zhou2025fidelitygpt,dramko2026idioms,li2026sk2decompile,zhang2026constraint,cui2026pcodetrans,liu2026binary,shypula2026decaf,gagnon2026practical}), and structural enrichment such as rodata, control-flow graphs (CFG), callees, .data, and .bss ($\text{I}_3$) \cite{cao2026behavior,zhou2025fidelitygpt,dramko2026idioms,cui2026pcodetrans,shypula2026decaf,gagnon2026practical,su2026rldecompiler}.
    \item \textbf{Augmentation (A):} The last axis depicts the inference-time mechanism divided into context provisioning through retrieval augmented generation (RAG, $\text{A}_1$) \cite{li2025pseudofix,zhou2025fidelitygpt,gagnon2026practical} and self-exemplar ($\text{A}_2$) \cite{feng2024self}), and verifier feedback, including compiler ($\text{A}_3$) \cite{wong2025decllm,cao2026behavior,zhang2026constraint,cui2026pcodetrans,liu2026binary,shypula2026decaf}, execution ($\text{A}_4$) \cite{wong2025decllm,cao2026behavior,zhang2026constraint,cui2026pcodetrans,liu2026binary}, and semantics ($\text{A}_5$) \cite{hu2024degpt}.
\end{enumerate}

\textbf{\textit{Research in 2024-2025:}} The field of LLM-assisted source code recovery opened with single-pass methods, wherein a single forward generation produces the recovered source. DeGPT~\cite{hu2024degpt} prompts an off-the-shelf GPT-3.5 with Ghidra pseudo-C passed through a fixed three-role pipeline comprising a referee that decides which of structure simplification, comment insertion, or variable renaming to apply, an advisor that proposes the edits, and an operator that accepts or rejects each edit through a Micro Snippet Semantic Calculation (MSSC) gate, adding no external data beyond that semantic check. LLM4Decompile~\cite{tan2024llm4decompile} fine-tunes DeepSeek-Coder into a decompiler series that translates in one pass under two input regimes, end-to-end from raw x86-64 assembly or refinement from Ghidra pseudo-C. SCCDEC~\cite{feng2024self} stays within single-pass control but adds self-augmentation: its tuning-free variant recompiles the model's own draft into a self-constructed (assembly, source) exemplar that is prepended before a second decompilation, and a companion fine-tuned variant further aligns the model to reach its best results. Nova~\cite{jiang2025nova} takes the most training-heavy route, pre-training an assembly-native model from DeepSeek-Coder with hierarchical attention and contrastive objectives and decompiling stripped x86-64 assembly in a single pass. PtxDec~\cite{sun2025enhancing} fine-tunes Qwen2.5-Coder-7B and Qwen3-32B to translate optimized GPU PTX, normalized into a compact rolled-PTX intermediate representation, back into readable CUDA. PseudoFix~\cite{li2025pseudofix} prompts off-the-shelf models with decompiler pseudo-C in a fixed detect-then-refactor sweep across six structural-distortion types, retrieving similar examples from an external distortion database to ground each fix. The lone departure from single-pass control was DecLLM~\cite{wong2025decllm}, which wraps off-the-shelf GPT-3.5 and GPT-4 in an iterative repair loop that recompiles and re-runs each candidate to repair the input pseudo-C, feeding back compiler errors, ASan reports, and test- and fuzzing-detected output divergences for up to fifteen rounds, foreshadowing the direction the field would take a year later.

\textbf{\textit{Research in 2026:}} Research in 2026 brought both a diversification of single-pass designs and a decisive shift toward iterative control. The sole fully agentic system (i.e. featuring an LLM with tool-calling capabilities), Cao~\cite{cao2026behavior}, deploys an off-the-shelf coding agent that autonomously reads AVR firmware disassembly and its interrupt-vector table and refines its output against avr-gcc recompilation and cycle-accurate simavr behavioral traces. Single-pass generation continued but grew richer in inputs and roles: FidelityGPT~\cite{zhou2025fidelitygpt} runs a two-role detect-then-correct pass over pseudo-C enriched with a program-dependence graph and retrieves from an annotated distortion database. Idioms~\cite{dramko2026idioms} fine-tunes local models to generate code and its user-defined type definitions jointly, feeding decompiler pseudo-C together with neighboring functions drawn from the call graph. Abualazm et al.~\cite{abualazm2026llms} fine-tune reasoning models to recover idiomatic Dart from the x86-64 assembly of ahead-of-time-compiled binaries. SK2Decompile~\cite{li2026sk2decompile} splits the task across two models, each supervised-fine-tuned then reinforcement-learning-tuned, a structure-recovery phase over IDA-Pro pseudo-C followed by an identifier-naming phase. RlDecompiler~\cite{su2026rldecompiler} trains an end-to-end model with supervised fine-tuning followed by Group Relative Policy Optimization (GRPO), whose input assembly is enriched with extracted string and floating-point constants and relabeled basic blocks. Finally, Decaf~\cite{shypula2026decaf} forgoes a loop for best-of-$N$ search, sampling many decompilations of Ghidra pseudo-C with declared types and globals, discarding those that fail to compile, and selecting the best with a learned re-ranker. The defining trend, however, was a wave of iterative feedback methods: Agent4Decompile~\cite{zhang2026constraint} drives three specialized repair LLM roles covering syntax, compilation, and execution from a scripted validator over decompiler pseudo-C. Although its name suggests an agentic control-flow, the models in this work do not exercise tool-calling agentic capability, making it an iterative, multi-role approach. PCodeTrans~\cite{cui2026pcodetrans} passes Hex-Rays pseudo-C plus a recovered type-dependency graph, globals, and callees through bounded compiler, sanitizer, and breakpoint-differential repair stages. AutoDecompiler~\cite{liu2026binary} fine-tunes with supervised then reinforcement learning to perform up-to-three-turn refinement of assembly- or pseudo-C-derived output against compiler, runtime, and I/O-test feedback. Finally, Gagnon et al.~\cite{gagnon2026practical} recover source by anchor-based retrieval instead of generation, extracting strings, constants, calls, and function names to query a source corpus and having the model verify and re-rank candidates while propagating confident matches through the call graph. Four of the five iterative methods are 2026 arXiv preprints, a clear pivot from one-shot generation toward compile- and execution-guided refinement.
\begin{table*}[t]
\centering
\caption{Evaluation metrics across the 22 papers (18 methods and 4 evaluations). $|P|$ and $|N|$ represent the number of papers and unique metrics per category, respectively. $\dagger$ indicate composite metrics.}
\label{tab:metrics}
\footnotesize
\setlength{\tabcolsep}{4pt}
\renewcommand{\arraystretch}{1.25}
\begin{tabular*}{\textwidth}{|l|l|c|c|>{\raggedright\arraybackslash}p{13.09cm}|}
\hline
\textbf{Axis} & \textbf{Objective} & \textbf{$|P|$} & \textbf{$|N|$} & \textbf{Unique Metrics} \\ \hline
Compilation & Quality & 13 & 5 & Re-compilation Rate~\cite{feng2024self,wong2025decllm,sun2025enhancing,cao2026behavior,su2026rldecompiler,zhang2026constraint,cui2026pcodetrans,liu2026binary,shypula2026decaf,gao2025decompilebench,yoda2026statisticalanalysisexecutabilityprogram}, Compile-error Distribution~\cite{sun2025enhancing,su2026rldecompiler,gao2025decompilebench}, Compile-error Repair Rate~\cite{wong2025decllm,cao2026behavior}, Compile@$k$~\cite{abualazm2026llms,abualazm2026evaluating}, Parse-only Syntax Rate~\cite{zhang2026constraint} \\ \hline
Readability & Quality & 9 & 12 & R2I$^\dagger$~\cite{li2026sk2decompile,liu2026binary,tian2026decompile}, User Study$^\dagger$~\cite{hu2024degpt,li2025pseudofix,zhou2025fidelitygpt}, LLM-judge Elo Rating$^\dagger$~\cite{su2026rldecompiler,gao2025decompilebench}, Halstead Complexity$^\dagger$~\cite{hu2024degpt,li2025pseudofix}, Comment Correctness and Non-triviality~\cite{hu2024degpt}, Cyclomatic Complexity~\cite{li2025pseudofix}, Goto Count~\cite{li2025pseudofix}, GPT-4o Readability Rating$^\dagger$~\cite{tan2024llm4decompile}, GPT-Judge Identifier Rating$^\dagger$~\cite{li2026sk2decompile}, LOC~\cite{li2025pseudofix}, Naming Coverage Ratio~\cite{hu2024degpt}, Nesting Depth~\cite{li2025pseudofix} \\ \hline
Lexical & Similarity & 11 & 8 & CodeBLEU$^\dagger$~\cite{sun2025enhancing,abualazm2026llms,tian2026decompile,abualazm2026evaluating,yoda2026statisticalanalysisexecutabilityprogram}, Edit Similarity~\cite{tan2024llm4decompile,shypula2026decaf,tian2026decompile}, BLEU-4~\cite{li2025pseudofix}, LOC Ratio~\cite{cao2026behavior}, Meaningful Variable Ratio~\cite{hu2024degpt}, METEOR~\cite{li2025pseudofix}, ROUGE-L~\cite{li2025pseudofix}, Variable Name Accuracy~\cite{dramko2026idioms} \\ \hline
Syntactic & Similarity & 5 & 1 & CodeBLEU$^\dagger$~\cite{sun2025enhancing,abualazm2026llms,tian2026decompile,abualazm2026evaluating,yoda2026statisticalanalysisexecutabilityprogram} \\ \hline
Semantic & Similarity & 17 & 19 & CodeBLEU$^\dagger$~\cite{sun2025enhancing,abualazm2026llms,tian2026decompile,abualazm2026evaluating,yoda2026statisticalanalysisexecutabilityprogram}, Dependency Equivalence~\cite{dramko2026idioms}, LLM-judge Elo Rating$^\dagger$~\cite{su2026rldecompiler,gao2025decompilebench}, GPT-4o Readability Rating$^\dagger$~\cite{tan2024llm4decompile}, GPT-Judge Identifier Rating$^\dagger$~\cite{li2026sk2decompile}, CodeQL Vulnerability Detection~\cite{wong2025decllm,shypula2026decaf}, Embedding Cosine Similarity~\cite{li2025pseudofix,tian2026decompile}, AVR Semantic Rubric$^\dagger$~\cite{cao2026behavior}, CodeScore~\cite{yoda2026statisticalanalysisexecutabilityprogram}, Distortion Fix Rate~\cite{zhou2025fidelitygpt}, GPT-Judge Recovery Scores$^\dagger$~\cite{tian2026decompile}, MSSC Acceptance~\cite{hu2024degpt}, Byte-wise Match Rate~\cite{shypula2026decaf}, Structure-fidelity Correctness~\cite{li2025pseudofix}, User-defined Type Accuracy~\cite{dramko2026idioms}, Variable Type Accuracy~\cite{dramko2026idioms}, Function-name Retention Rate~\cite{zhang2026constraint}, Vulnerability Pattern Preservation~\cite{zhang2026constraint}, 9-dimension Quality Score$^\dagger$~\cite{yoda2026statisticalanalysisexecutabilityprogram} \\ \hline
Functional & Similarity & 16 & 13 & Re-executability Rate~\cite{tan2024llm4decompile,feng2024self,wong2025decllm,dramko2026idioms,li2026sk2decompile,su2026rldecompiler,zhang2026constraint,cui2026pcodetrans,liu2026binary,shypula2026decaf,tian2026decompile}, Pass@$k$~\cite{jiang2025nova,su2026rldecompiler,abualazm2026evaluating}, ASan Repair Rate~\cite{wong2025decllm,cui2026pcodetrans}, Average Fuzzing Coverage~\cite{zhang2026constraint}, Breakpoint-differential Repair Rate~\cite{cui2026pcodetrans}, Coverage Equivalence Rate~\cite{gao2025decompilebench}, Exploit Trigger Rate~\cite{zhang2026constraint}, Fuzz-ready Rate~\cite{zhang2026constraint}, Fuzzing-detected Divergence Count~\cite{wong2025decllm}, I/O Correctness Rate~\cite{sun2025enhancing}, Runtime Correctness Rate~\cite{sun2025enhancing}, Scheduling Consistency$^\dagger$~\cite{sun2025enhancing}, Trace Equivalence Rate~\cite{cao2026behavior} \\ \hline
\end{tabular*}
\end{table*}

\begin{figure}
    \centering
    \includegraphics[width=\linewidth]{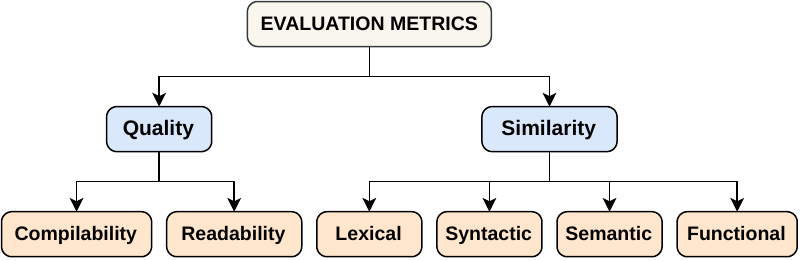}
    \caption{The taxonomy of evaluation metrics for LLM-assisted source recovery.}
    \label{fig:metrics}
\end{figure}

\section{Evaluation Metrics Used in Source Recovery}
\label{sec:metrics}

Evaluation metrics are a key methodological decision to ensure the quality of source recovery \cite{liu2026llm}. Intuitively, a recovered source code must be compilable and easy to read (quality), and it should mean and behavior similarly to original source code while also retaining its appearance and structure (similarity). In this regard, we suggest a holistic categorization of evaluation metrics for source recovery into the two objectives and six dimensions, as depicted in Fig.~\ref{fig:metrics}. The dimensions are defined as follows, citing papers from source-code recovery that include the dimensions in their evaluations:
\begin{enumerate}
    \item \textbf{Compilation:} Depicts whether the recovered code parses, compiles, and links \cite{feng2024self,wong2025decllm,sun2025enhancing,cao2026behavior,abualazm2026llms,su2026rldecompiler,zhang2026constraint,cui2026pcodetrans,liu2026binary,shypula2026decaf,gao2025decompilebench,abualazm2026evaluating,yoda2026statisticalanalysisexecutabilityprogram}.
    \item \textbf{Readability:} Depicts how easy the generated code is to read and is evaluated using complexity indices, naming and comment quality, human ratings, or LLM judges \cite{hu2024degpt,tan2024llm4decompile,li2025pseudofix,zhou2025fidelitygpt,li2026sk2decompile,su2026rldecompiler,liu2026binary,gao2025decompilebench,tian2026decompile}.
    \item \textbf{Lexical:} Compares the generated code to the original source at the text and token level \cite{hu2024degpt,tan2024llm4decompile,sun2025enhancing,li2025pseudofix,cao2026behavior,dramko2026idioms,abualazm2026llms,shypula2026decaf,tian2026decompile,abualazm2026evaluating,yoda2026statisticalanalysisexecutabilityprogram}.
    \item \textbf{Syntactic:} Compares the Abstract Syntax Tree (AST) structure to the original \cite{sun2025enhancing,abualazm2026llms,tian2026decompile,abualazm2026evaluating,yoda2026statisticalanalysisexecutabilityprogram}.
    \item \textbf{Semantic:} Compares meaning of the generated and original code, established statically over code, graphs, or recompiled binaries without code execution \cite{hu2024degpt,tan2024llm4decompile,wong2025decllm,sun2025enhancing,li2025pseudofix,cao2026behavior,zhou2025fidelitygpt,dramko2026idioms,abualazm2026llms,li2026sk2decompile,su2026rldecompiler,zhang2026constraint,shypula2026decaf,gao2025decompilebench,tian2026decompile,abualazm2026evaluating,yoda2026statisticalanalysisexecutabilityprogram}.
    \item \textbf{Functional:} Matches runtime behavior, established by executing the recompiled code \cite{tan2024llm4decompile,feng2024self,wong2025decllm,sun2025enhancing,jiang2025nova,cao2026behavior,dramko2026idioms,li2026sk2decompile,su2026rldecompiler,zhang2026constraint,cui2026pcodetrans,liu2026binary,shypula2026decaf,gao2025decompilebench,tian2026decompile,abualazm2026evaluating}.
\end{enumerate}

\emph{Compilation} and \emph{Readability} measure the general ``quality" of the generated code, while \emph{Lexical}, \emph{Syntactic}, \emph{Semantic}, and \emph{Functional} similarity measure how "close" or "equivalent" (if derivable) the generated source code is to the original in terms of its appearance, structure, meaning, and behavior, respectively. The distinction between the two objectives is the comparison target: quality metrics are computed on the generated code alone, or against the decompiler output a method refines, whereas similarity metrics are computed against the original program. Table \ref{tab:metrics} compiles $53$ unique evaluation metrics extracted from the $18$ method and $4$ evaluation papers on LLM-assisted source recovery. $11$ of the $53$ metrics, marked~$\dagger$ in Table~\ref{tab:metrics}, are composites that fold several sub-dimensions into a single score or rating protocol. We count each composite as one metric, and a composite whose sub-dimensions span multiple dimensions is listed on every axis it measures rather than split into constituent scores.

\subsubsection{Code Quality Metrics}
Among the quality metrics, prior works on LLM-assisted source recovery use $5$ unique metrics for \textbf{compilation}: nearly every paper measures a plain \emph{Re-compilation Rate (RC)}~\cite{feng2024self,wong2025decllm,sun2025enhancing,cao2026behavior,su2026rldecompiler,zhang2026constraint,cui2026pcodetrans,liu2026binary,shypula2026decaf,gao2025decompilebench,yoda2026statisticalanalysisexecutabilityprogram}, and the rest of the variation is limited to best-of-$k$ compilation as \emph{compile@$k$}~\cite{abualazm2026llms,abualazm2026evaluating}, \emph{parse-only syntax checks}~\cite{zhang2026constraint}, and the \emph{repair}~\cite{wong2025decllm,cao2026behavior} and \emph{distribution}~\cite{sun2025enhancing,su2026rldecompiler,gao2025decompilebench} of compile errors. Since compilation is also the cheapest signal to obtain, it doubles as the feedback channel of the iterative methods discussed above. \textbf{Readability} is more fragmented, with $12$ metrics across $9$ papers. These include the Relative Readability Index (\emph{R2I})~\cite{li2026sk2decompile,liu2026binary,tian2026decompile}, classical complexity indices, namely \emph{Halstead}, \emph{cyclomatic complexity}, \emph{nesting depth}, Lines of Code (\emph{LOC}), and \emph{goto count}~\cite{hu2024degpt,li2025pseudofix}, naming and comment measures~\cite{hu2024degpt}, $3$ LLM-judge protocols~\cite{tan2024llm4decompile,li2026sk2decompile,su2026rldecompiler,gao2025decompilebench}, and $4$ user studies~\cite{hu2024degpt,li2025pseudofix,zhou2025fidelitygpt} whose participant pools, questions, and rating scales all differ, which we therefore club into a single \emph{user study} category.

\subsubsection{Code Similarity Metrics}
The identified prior works measure \textbf{lexical similarity} using $8$ metrics, dominated by CodeBLEU's \emph{n-gram} component~\cite{sun2025enhancing,abualazm2026llms,tian2026decompile,abualazm2026evaluating,yoda2026statisticalanalysisexecutabilityprogram}, and \emph{edit similarity}~\cite{tan2024llm4decompile,shypula2026decaf,tian2026decompile}. \emph{BLEU-4}, \emph{METEOR}, and \emph{ROUGE-L} appear in a single paper~\cite{li2025pseudofix}, and a \emph{LOC ratio} serves as a coarse size-fidelity metric~\cite{cao2026behavior}. Identifier recovery is also scored at the token level against the original identifiers, by \emph{variable name accuracy}~\cite{dramko2026idioms} and the \emph{meaningful variable ratio}~\cite{hu2024degpt}; both are string comparisons and, unlike the meaning-level instruments below, neither is invariant to renaming. \textbf{Syntactic similarity} exists in this literature only as CodeBLEU's \emph{AST-match} component \cite{sun2025enhancing,abualazm2026llms,tian2026decompile,abualazm2026evaluating,yoda2026statisticalanalysisexecutabilityprogram}. \textbf{Semantic similarity} is the broadest and least standardized axis, with $19$ metrics across $17$ papers. It spans metrics for static graph comparison, including \emph{dependency equivalence}~\cite{dramko2026idioms}, which tests dependency-graph isomorphism as an approximation of program equivalence, and CodeBLEU's \emph{Data Flow (DF)-match}, and recovery accuracy for \emph{variable types}, \emph{user-defined types}~\cite{dramko2026idioms}, and \emph{function names}~\cite{zhang2026constraint}. Binary-level equivalence is established by the \emph{byte-wise match} of recompiled functions~\cite{shypula2026decaf}, and \emph{structure-fidelity correctness} is annotated manually against the original source~\cite{li2025pseudofix}. Learned similarity metrics include \emph{embedding cosine similarity}~\cite{li2025pseudofix,tian2026decompile} and \emph{CodeScore}~\cite{yoda2026statisticalanalysisexecutabilityprogram}, and downstream utility is evaluated with \emph{CodeQL vulnerability detection}~\cite{wong2025decllm,shypula2026decaf} and \emph{vulnerability pattern preservation}~\cite{zhang2026constraint}. Judged and rubric-based composites, namely those of Cao~\cite{cao2026behavior} and Yoda et al.~\cite{yoda2026statisticalanalysisexecutabilityprogram}, the correctness half of the LLM-judge Elo criteria~\cite{gao2025decompilebench,su2026rldecompiler}, and the source-referencing GPT judges~\cite{tan2024llm4decompile,li2026sk2decompile,tian2026decompile}, cut across these semantic families. \textbf{Functional similarity} is evaluated in $16$ of the $22$ papers by executing the recompiled code: $11$ of them reporting \emph{re-executability}~\cite{tan2024llm4decompile,feng2024self,wong2025decllm,dramko2026idioms,li2026sk2decompile,su2026rldecompiler,zhang2026constraint,cui2026pcodetrans,liu2026binary,shypula2026decaf,tian2026decompile} and $3$ reporting \emph{pass@$k$}~\cite{jiang2025nova,su2026rldecompiler,abualazm2026evaluating}, while the rest test runtime behavior through \emph{trace equivalence}~\cite{cao2026behavior}, \emph{coverage equivalence}~\cite{gao2025decompilebench}, sanitizer-guided repair~\cite{wong2025decllm,cui2026pcodetrans}, and fuzzing~\cite{wong2025decllm,zhang2026constraint}.

\section{Evaluation Methodology}
\label{sec:methodology}
This section describes the evaluation metrics, selected SOTA methods and models, experimental design, and the dataset used in our empirical evaluations. 

\subsection{Evaluation Metrics Suite \colorbox{red!20}{\texttt{RQ1}}}
\label{subsec:metrics}
Table~\ref{tab:metrics} highlighted the fragmentation of evaluation metrics across research papers on LLM-assisted source recovery. None of the surveyed evaluation and method papers include an evaluation suite that captures the six dimensions along which recovered source can fail. To conduct a holistic and interpretable evaluation, we adopt one principled metric per dimension, with the exception of semantic similarity which has two metrics to cover source-derived and binary-derived semantics, resulting in a 7 metric suite described below. 

Re-compilation rate (RC) and Re-execution Rate (RE) are the most prevalent compilation quality and functional similarity metrics reported in prior works. In its traditional form, $RC$ is the ratio of generated functions that compile (standalone) to the total number of evaluation functions. Our results (Fig. \ref{fig:sota_funnel} and \ref{fig:representation_funnel}) indicate that LLMs are better at generating compilable code than recovering the correct signature. This takeaway is typically missed by RC and muddled in RE which targets functional correctness. Thus, Re-compilation Rate with caller Interface (RCI) is a valuable metric that can be computed using a benchmark's evaluation harness. Further, RC, RCI, and RE follow a nested relationship ($RE\leq RCI \leq RC$): \emph{RCI} and \emph{RE} are our choice for compilation quality and functional similarity metrics, respectively. 

R2I \cite{eom2024r2i} provides 31-readability feature composition and is increasingly being adopted as a readability metric. To standardize, we define $\Delta_\text{R2I}$ as our readability quality metric to the relative R2I between a generated code $\hat{c}$ and the pseudo-C of the original candidate $c_\text{pc}$ as follows:
\begin{equation}
\Delta_{\text{R2I}}(\hat{c},c_\text{pc}) = \text{R2I}(\hat{c}) - \text{R2I}(c_{\text{pc}}).
\label{eq:r2i}
\end{equation}
A positive $\Delta_{\text{R2I}}$ means the model's generation is more readable than the decompiler pseudo-C an analyst would otherwise work with. We compare with the standardized decompiler output for a standardized relative quality measure instead of with noisy source code.  

We use \emph{Edit Similarity (ES)} for lexical similarity. ES measures surface fidelity by computing the normalized Levenshtein distance between two code snippets reported in the range $[0,1]$, with higher values indicating closer surface form. We use CodeBLEU's \emph{AST-match (AST)} \cite{shuo2009codebleu} to abstract-syntax-tree structure for syntactic similarity. AST is largely insensitive to the naming noise that dominates ES and provides a more specific measure of the relative syntactical structure of the recovered source. It is also the only syntactic similarity metric used in the surveyed literature in Section \ref{sec:systematization}. CodeBLEU's \emph{DF-match (DF)} \cite{shuo2009codebleu} is used for semantic similarity using source-derived DF comparison: it matches def-then-use graphs, abstracting away both naming and syntactic idiom to test whether values flow through the recovered code as they do in the original. These three metrics compare against the original source. \emph{Byte-wise match (BW)} \cite{shypula2026decaf} instead provides semantic similarity at the binary level, recompiling the generated function with the original compiler and flags and comparing the resulting machine code byte-for-byte against the original binary (wildcarding relocation sites). BW is the strictest metric in our suite since any divergence in instruction selection fails it, even between behaviorally identical programs and a pass certifies recovery up to compiler determinism.

\subsection{Experiments and Models}
\label{sec:experiments}

We conduct three empirical evaluations, detailed as follows: \textbf{First,} ``SOTA Method Evaluation" (reported in Section \ref{subsec:sota_eval}) evaluates the impact of the following: \textbf{(1)} Design choices of SOTA methods following the taxonomy defined in Section \ref{sec:systematization}; \textbf{(2)} Generalization of SOTA methods over four standard (x86-64, aarch64, riscv64, x86-32) and five embedded (armv7m, armv8m, armv7a, armv8a32, mips32le) architectures, five optimization levels (\texttt{O0-O3, Os}), and stripped and unstripped symbol variants; and \textbf{(3)} Spearman correlation of the metrics suite derived from the above. We select LLM4Decompile (assembly and Ghidra)~\cite{tan2024llm4decompile}, SCCDEC~\cite{feng2024self}, Nova~\cite{jiang2025nova}, SK2Decompile~\cite{li2026sk2decompile}, DeGPT~\cite{hu2024degpt}, FidelityGPT~\cite{zhou2025fidelitygpt}, and Agent4Decompile~\cite{zhang2026constraint} for evaluations, encompassing majority of the identified methodology axes ($\text{M},\text{C},\text{R},\text{I},\text{A}$) and their subcategories, subject to code availability. Only $\text{C}_3$ (agentic control-flow) has no representative method in our evaluation since Cao~\cite{cao2026behavior}, which is the only identified agentic approach, has not published its code. We note again that Agent4Decompile~\cite{zhang2026constraint} is classified as an iterative method. Methods that fine-tune or pre-train their own models are evaluated using their specified models, while methods that use off-the-shelf models are  evaluated using HuggingFace sourced \texttt{gemma4-31b-it} \cite{gemmateam2026gemma4} for comparison at the same intelligence capability.

\textbf{Second,} ``Taxonomy-driven Pipeline Ablation" (reported in Section \ref{subsec:taxonomy_ablation}) ablates five aspects of the source recovery pipeline: \textbf{(1)} Base input representation  (hex, assembly, IR, pseudo-C) provided to a single-role, single-pass pipeline (SP), and subsequently; \textbf{(2)} Enrichment channels (callee imports, .rodata, .data/.bss, CFG, and frame ABI) that aid SP with additional context; \textbf{(3)} LLMs of different scale in the SP pipeline; \textbf{(4)} Iterative control-flow with one compiler diagnostic feedback iteration (CIT); and \textbf{(5)} Review-role (RR) pipeline using a reviewer LLM to improve the code based on static analysis. Each pipeline uses open-weight LLMs from HuggingFace: \texttt{gemma4-12b-it} \cite{gemmateam2026gemma4}, \texttt{gemma4-31b-it} \cite{gemmateam2026gemma4} and \texttt{qwen3.8-27b} \cite{qwen38} at \texttt{bf16} quantization, 0.1 temperature, and a 8192-token generation limit and within a 32768-token context. For reference, a CIT or RR pipeline with a 31b LLM is represented as ``CIT-31b" or ``RR-31b", respectively. Ablations are conducted on 50 C-code compiled on x86-64 at two optimizations (O0 and O2). 

\textbf{Third,} ``Source Language Study" (reported in Section \ref{subsec:language_study}) uses the representation selected by RQ3 to test \textbf{(1)} same-language recovery, where the model is told which language (among C, C++, Go, Rust, Fortran and Ada) the binary was built from and must reproduce it, and \textbf{(2)} cross-language recovery, where it must produce behaviorally equivalent C from a non-C-derived binary. These tasks are conducted using the SP and CIT pipelines, and \texttt{gemma4-31b-it}  and \texttt{qwen3.8-27b}. C is selected as the cross-language target as it has the most mature tooling for downstream vulnerability and malware analysis.

\subsection{Dataset}
\label{sec:dataset}
We assemble our dataset using the \texttt{test\_real} split of Exebench~\cite{armengol2022exebench} that originally comprises of 2,134 C-language functions, filtered down to 1,324 to ensure compilability and reference I/O case passes on the target architectures and optimizations in this study. We stratify the reduced functions by Halstead difficulty~\cite{halstead1977elements, bar2026-does-representation-matter} defined as follows:
\begin{equation}
D = \frac{\eta_1}{2}\cdot\frac{N_2}{\eta_2},
\label{eq:halstead}
\end{equation}
where $\eta_1$ is the number of \emph{distinct} operators in the function, $\eta_2$ is the number of \emph{distinct} operands, and $N_2$ is the \emph{total} number of operand occurrences. The ratio $N_2/\eta_2$ is therefore the average number of times an operand is reused, so that $D$ grows with both operator variety and operand reuse, which is the notion of difficulty we want the sample to span. We then select $500$ functions by rank-spacing over the \emph{distinct} difficulty values, which avoids both the duplicate mass of plain rank sampling and the outlier domination of value spacing. Fig. \ref{fig:halstead} (Appendix) highlights the difficulty distribution in the filtered split and selected functions across difficulty groups, highlighting a right-skewed set with a long (but sparse) tail due to the available distribution in the original split.

For answering RQ2, we use $500$ functions across nine architectures (86-64, aarch64, riscv64, x86-32, armv7m, armv8m, armv7a, armv8a32, mips32le), five optimization levels (\texttt{O0-O3, Os}), and stripped and unstripped variants, resulting in \textbf{45,000 test samples}. For RQ3, and RQ4,  we take a $50$-function subset sampled uniformly from the $500$-function corpus compiled with the x86-64 gcc toolchain at \texttt{O0} and \texttt{O2} optimizations and stripped. The experiments for RQ4 use a set of $300$ functions, comprising $50$ functions each in C, C++, Go, Rust, Fortran and Ada, wherein the non-C functions are translations of the C functions done using Claude Fable-5.1 \cite{anthropic_fable_2026}. The translations were validated against the I/O test cases of the C functions to ensure functional similarity and checked by the authors.

\section{Evaluations and Discussion}
\label{sec:results}
This section presents the empirical results of the experiments described in Section \ref{sec:experiments} and their implications. We summarize our findings in Section~\ref{sec:key-findings-OP}.

\begin{figure}
    \centering
    \includegraphics[width=\linewidth]{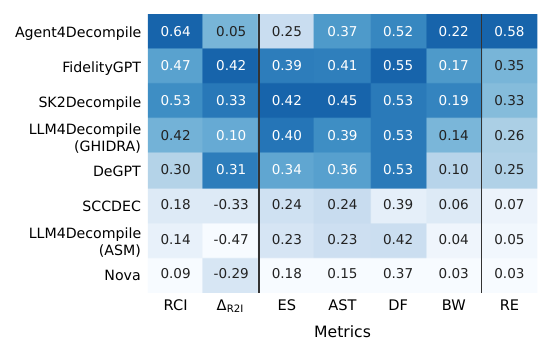}
    \caption{SOTA methods scored on all metrics averaged across architectures, optimizations, and symbol variants}
    \label{fig:sota_metrics}
\end{figure}

\begin{figure}
    \centering
    \includegraphics[width=0.9\linewidth]{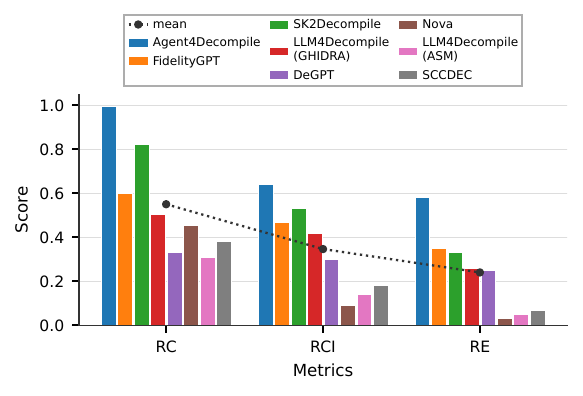}
    \caption{RC, RCI, and RE per method.}
    \label{fig:sota_funnel}
\end{figure}

\begin{figure}
    \centering
    \includegraphics[width=\linewidth]{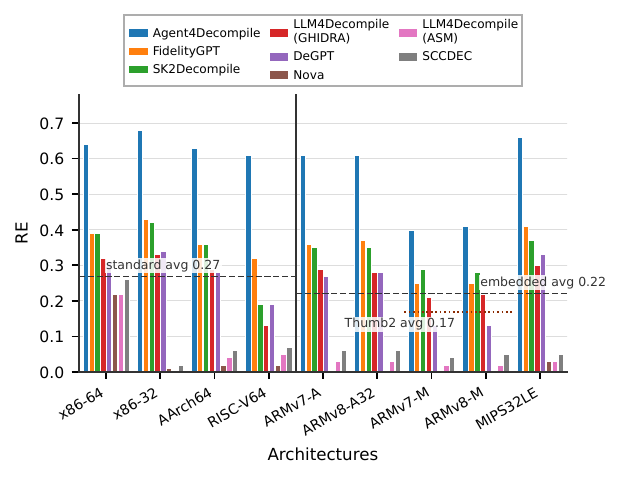}
    \caption{RE of methods across standard and embedded architectures, averaged over optimizations and symbol variants.}
    \label{fig:sota_architectures}
\end{figure}

\begin{figure}
    \centering
    \includegraphics[width=\linewidth]{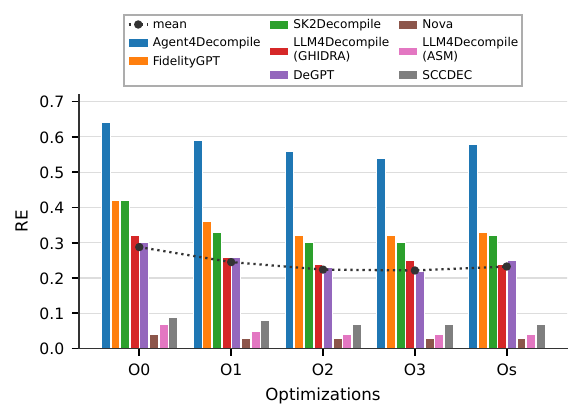}
    \caption{RE of methods across optimizations, averaged over architectures and symbol variants.}
    \label{fig:sota_optimizations}
\end{figure}

\begin{figure}
    \centering
    \includegraphics[width=\linewidth]{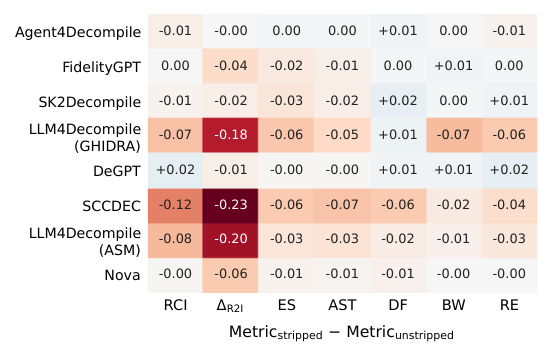}
    \caption{Difference between metrics in stripped and unstripped cases to highlight the impact of symbol stripping.}
    \label{fig:sota_symbols}
\end{figure}

\subsection{SOTA Method Evaluation \colorbox{red!20}{\texttt{RQ1-3}}}
\label{subsec:sota_eval}

Fig. \ref{fig:sota_metrics} scores the selected SOTA methods on all seven metrics (RCI, $\Delta_\text{R2I}$, ES, AST, DF, BW, RE), averaged over architectures, optimizations, and symbol variants. Fig.~\ref{fig:sota_funnel} highlights the comparative trend of RC, RCI and RE across SOTA approaches. Fig.~\ref{fig:sota_architectures} and~\ref{fig:sota_optimizations} present their average RE across architectures and optimizations, while Fig.~\ref{fig:sota_symbols} presents RE and $\Delta_\text{R2I}$ across stripped and unstripped cases. Tables \ref{tab:rq2_arch}, \ref{tab:rq2_opt} and \ref{tab:rq2_symbols} in the Appendix compile comprehensive empirical results, while Fig. \ref{fig:halstead_performance} depicts RE against Halstead groups.

\subsubsection{Impact of SOTA Design Choices}

We first present our observations of SOTA methods linked to the taxonomy described in Fig.~\ref{fig:taxonomy} and Table \ref{tab:sota}. As depicted in Fig. \ref{fig:sota_metrics}, methods that ingest pseudo-C (DeGPT, LLM4Decompile-Ghidra, FidelityGPT, SK2Decompile, and Agent4Decompile) outperform those that use assembly across all metrics, architectures, optimizations, and symbol variants, regardless of other harness differences: averaged over their tiers, pseudo-C methods reach RCI 0.47, $\Delta_\text{R2I}$ 0.24, ES 0.36, AST 0.40, DF 0.53, BW 0.17, and RE 0.36, against assembly-based methods. Further, methods that use off-the-shelf, general-purpose models (Agent4Decompile, FidelityGPT, DeGPT) achieve fidelity comparable to counterparts fine-tuned on pseudo-C (SK2Decompile, LLM4Decompile-Ghidra), and significantly better than those fine-tuned or pre-trained on source recovery from assembly (SCCDEC, LLM4Decompile-ASM, Nova). Interestingly, SK2Decompile is among the top half of methods when tested using Ghidra pseudo-C despite its IDA Pro pseudo-C fine-tuning.

Examining the control flow and multi-role methods, we note that iterative multi-role methods (Agent4Decompile) report the highest RCI, BW, and RE compared to all other control flows through their targeted refinement iterations using in-loop compiler and functional feedback \cite{zhang2026constraint}. However, Agent4Decompile's $\Delta_\text{R2I}$ (0.05) and ES (0.25) is significantly worse than non-iterative, multi-role methods (FidelityGPT:~0.42, 0.39; SK2Decompile: 0.33, 0.42; DeGPT: 0.31, 0.34) which report the highest $\Delta_\text{R2I}$ and ES, with similar DF and AST. LLM4Decompile-GHIDRA is the best-performing single-pass and single-role method, attributed to its pseudo-C training. Lastly, multi-role methods (Agent4Decompile, FidelityGPT, DeGPT, SK2Decompile) generally outperform single-role ones (LLM4Decompile-ASM, Nova, SCCDEC).

\subsubsection{Generalization Across Architectures, Optimizations, and Symbol Variants} 
As Fig.~\ref{fig:sota_architectures} shows, the RE of SOTA methods on embedded microcontroller architectures based on Thumb2 ISA (armv7-m, armv8m) is worse than standard (x86-64, x86-32, AArch64, RISC-V64) and higher-end embedded architectures (armv7-a, armv8-a32, MIPS32LE), highlighting a limitation of current methods and their LLMs in source recovery for IoT microcontrollers. Fig. \ref{fig:sota_optimizations} presents a monotonic drop in RE from \texttt{O0--O3} and similar performance between \texttt{O2} and \texttt{Os}. Fig. \ref{fig:sota_symbols} highlights that symbol stripping causes a minimal drop in performance across the top methods, including Agent4Decompile, FidelityGPT, SK2Decompile and DeGPT, while impacting $\Delta_\text{R2I}$ the most for other methods. 

\noindent{\bf Summary}: Overall, Agent4Decompile is the best performing method in terms of RCI, BW and RE across architectures and optimizations, showing a 41\% RE on Thumb2 architectures, a 43\% improvement compared to the next best method, i.e. SK2Decompile. It has comparable AST and DF with FidelityGPT and SK2Decompile, compared to which its limitation lies in readability and lexical similarity.  

\subsubsection{Metric Correlation}
\begin{figure}
    \centering
    \includegraphics[width=0.7\linewidth]{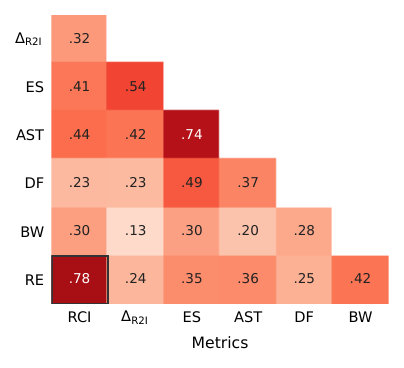}
    \caption{Metric correlation across 45,000 source generations.}
    \label{fig:metric_correlation}
\end{figure}

Fig. \ref{fig:metric_correlation} presents the Spearman correlation \cite{hauke2011comparison} between the 7 metrics evaluated from all 45,000 cases (500 functions $\times$ 9 architectures $\times$ 5 optimizations $\times$ 2 symbols). Each metric was picked based on its coverage of different failure modes in terms of quality and similarity, as described in Section \ref{sec:metrics}. The correlation figure shows that the selected metrics carry largely non-overlapping information. RE and RCI share the highest correlation (0.78) due to nesting, since RE can only be computed when a function compiles. However, they carry different information by definition. 

\begin{figure}[t]
    \begin{subfigure}[t]{\linewidth}
        \centering
        \includegraphics[width=\linewidth]{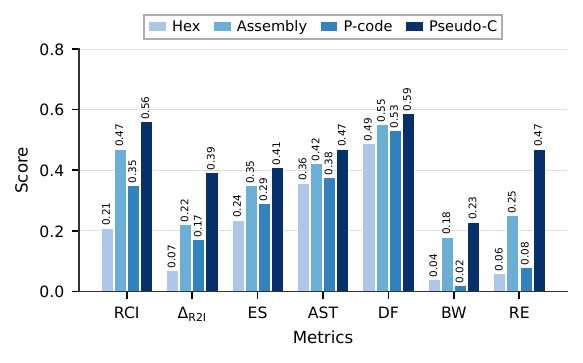}
        \caption{Base input representations.}
        \label{fig:rq3_input_representation}
    \end{subfigure}%
    \\
    \begin{subfigure}[t]{\linewidth}
        \centering
        \includegraphics[width=\linewidth]{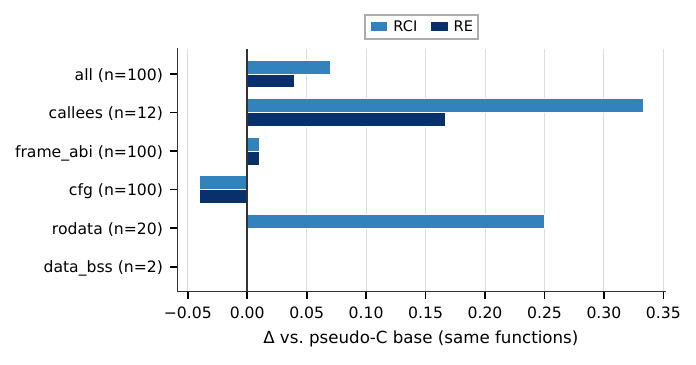}
        \caption{Contextual enrichment.}
        \label{fig:rq3_enrichment}
    \end{subfigure}%
    \centering
    \caption{Input representation ablations using SP-31b.}
    \label{fig:rq3_inputs}
\end{figure}

\subsection{Taxonomy-driven Pipeline Ablation \colorbox{red!20}{\texttt{RQ3}}}
\label{subsec:taxonomy_ablation}

We now report the results of evaluations based on targeted and controlled, taxonomy-driven ablation using our SP, CIT, and RR pipelines and three off-the-shelf LLMs.

\begin{figure}
    \centering
    \includegraphics[width=\linewidth]{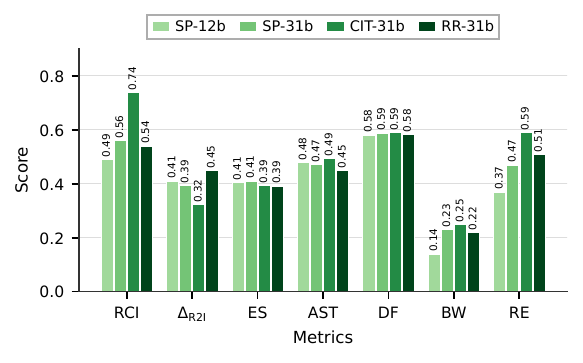}
    \caption{Impact of model scale in single-pass, iteration with compiler feedback, and additional LLM roles for code review.}
    \label{fig:rq3_model_iteration_role}
\end{figure}

\begin{figure}[t]
    \begin{subfigure}[t]{\linewidth}
        \centering
        \includegraphics[width=\linewidth]{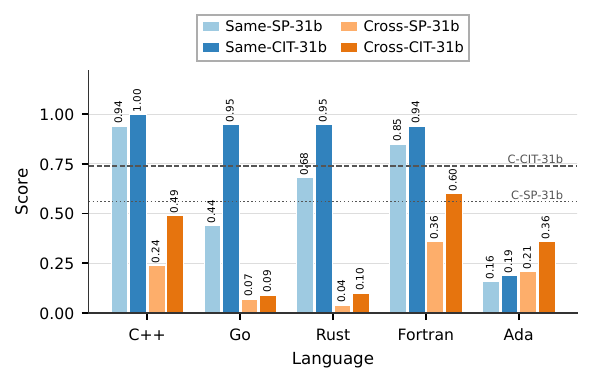}
        \caption{RCI}
        \label{fig:same_vs_cross_rci}
    \end{subfigure}%
    \\
    \begin{subfigure}[t]{\linewidth}
        \centering
        \includegraphics[width=\linewidth]{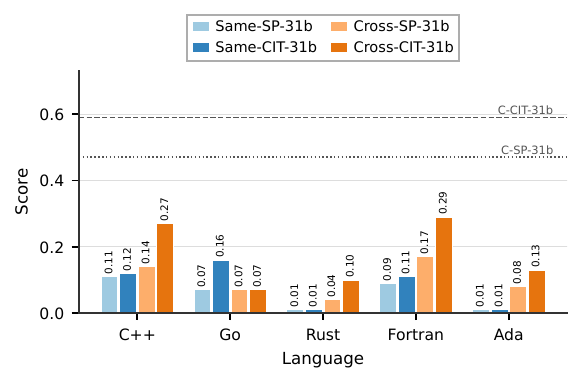}
        \caption{RE}
        \label{fig:same_vs_cross_re}
    \end{subfigure}%
    \centering
    \caption{RCI and RE of same- and cross-language recovery using SP-31b and CIT-31b pipelines.}
    \label{fig:rq4_languages}
\end{figure}

\begin{figure}[t]
    \begin{subfigure}[t]{\linewidth}
        \centering
        \includegraphics[width=0.85\linewidth]{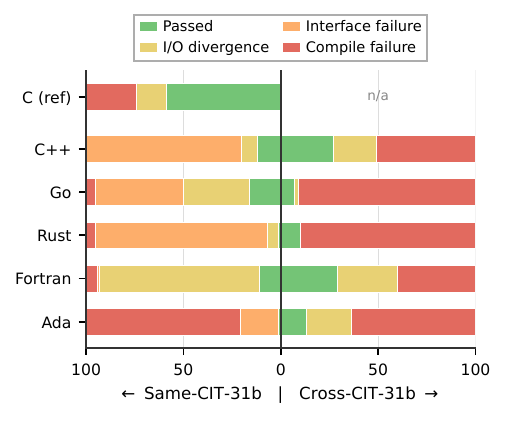}
        \caption{Failure stages across same- and cross- language recovery using CIT-31b pipelines.}
        \label{fig:same_vs_cross_stages}
    \end{subfigure}%
    \\
    \begin{subfigure}[t]{\linewidth}
        \centering
        \includegraphics[width=0.75\linewidth]{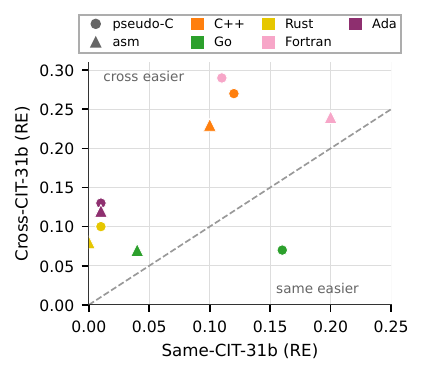}
        \caption{Comparison of RE in same-CIT-31b and cross-CIT-31b from \texttt{pseudo-C} (circle) and \texttt{asm} (triangle).}
        \label{fig:same_cross_inversion}
    \end{subfigure}%
    \centering
    \caption{Failure stages and same/cross comparison across languages.}
    \label{fig:rq4_language_inversion}
\end{figure}



\subsubsection{Input Representation}

Table~\ref{tab:rq3_ladder_enrichment} in Appendix presents the full set of results for \texttt{O0/O2} across 7 metrics derived from the three model scales. Fig.~\ref{fig:rq3_input_representation} depicts the ablation of \texttt{hex, asm, Ghidra p-code} and \texttt{Ghidra pseudo-C} using the SP-31b pipeline on 100 samples covering 50 functions compiled on x86 with O0 and O2 optimizations with stripped symbols. Evidently, pseudo-C achieves the best relative performance across all metrics. \texttt{gemma4-12b-it} and \texttt{qwen3.8-27b} show the same relative pattern, as seen in Table \ref{tab:rq3_ladder_enrichment} (Appendix). Further, Fig. \ref{fig:rq3_enrichment} presents the impact of various enrichment channels, including callee imports, .rodata, .data/.bss, CFG, frame ABI and all (combined enrichment). Frame ABI and cfg are available for all samples, but show limited or negative improvement in RCI and RE, respectively. Callees and .rodata are available in 12\% and 20\% of the samples, respectively and show over 20\% improvement. Overall, the result is inconclusive due to minimal improvement or limited availability of enrichment channels. We select pseudo-C as the base representation for further analysis.

\subsubsection{Model Scale, Iterations, and Review Roles}

Fig.~\ref{fig:rq3_model_iteration_role} presents the per-metric results of four source recovery harnesses, namely, SP-12b, SP-31b, CIT-31b, and RR-31b. CIT-31b and RR-31b reuse the generated source C and the Ghidra pseudo-C derived from the original binary for a second pass: CIT-31b includes compiler feedback to improve the generation, while RR-31b is only provided with instructions for static analysis without compiler feedback. 

We observe 14.3\%, 64\%, and 37\% improvement in RCI, BW, and RE when switching SP-12b to SP-31b, showing that among recent general-purpose models, bigger LLMs achieve better performance. A change from SP-12b to SP-27b (\texttt{qwen3.8-27b}) shows a similar pattern and is reported in the Appendix. Further, we observe improvements of 32\% RCI, 5\% AST, 8.7\% BW, and 25.5\% RE when switching SP-31b to CIT-31b, indicating the impact of targeted, high-quality functional feedback provided by the compiler diagnostics in helping the LLM fix its previous generation. However, CIT-31b incurs a degradation of 18.3\% $\Delta_\text{R2I}$ and 3.8\% ES due to defensive code frames added into new generations for compilability. This also explains the poor readability and ES observed in Agent4Decompile compared to FidelityGPT and SK2Decompile. Moreover, 25.5\% RE and 8.7\% BW gains through pure compiler feedback highlights the functional and semantic similarity hidden behind compilation errors. Finally, the reviewer role in RR-31b improves $\Delta_\text{R2I}$ by 13.8\% and RE by 8.5\% over SP-31b, with a 3.6-4.6\% reduction in every other metric. Table \ref{tab:rq3_scale_iteration} in the Appendix depicts optimization-wise metrics for the four tested pipelines.

\subsection{Source Language Study \colorbox{red!20}{\texttt{RQ4}}}
\label{subsec:language_study}

Fig.~\ref{fig:rq4_languages} presents the RCI and RE of same- and cross-language recovery using SP-31b and CIT-31b pipelines for C, C++, Go, Rust, Fortran, and Ada. Same-language recovery exposes a performance gap when viewed through the metrics of compilability and functional similarity. With the exception of Ada (16\% RCI), non-C languages recompile at rates comparable to or better than C (C++ 94\%, Fortran 85\%, Rust 68\%), yet their RE collapses to 1-13\% compared to 47\% for C. A single compiler iteration widens this split: CIT-31b drives RCI up to 94--100\% C++, Go, Rust, and Fortran (94-100\%) while RE improves from 7\% 16\% for Go and only by 1\% for Rust. The failure-stage decomposition in Fig.~\ref{fig:same_vs_cross_stages} shows that 80--88\% of C++ and Rust cases compile but cannot be \emph{called} by the evaluation harness due to incorrect signatures. Fortran instead runs and diverges (82\% I/O divergence) and Ada does not benefit from compilation repair.

Cross-language cases present an interesting observation. Single-pass RCI is far lower (4--36\%) because the model must strip the source language's runtime scaffolding from the pseudo-C. That being the case, successful compilations overwhelmingly pass the I/O tests. Further, iteration repairs leads to nearly 2$\times$ functional gains for Fortran (17\%$\to$29\%), C++ (14\%$\to$27\%), and Ada (8\%$\to$13\%). Go and Rust see poor performance since their binaries embed goroutine scheduling, bounds-check panics, and garbage-collection machinery that the LLM is unable to translate into plain C. Fig.~\ref{fig:same_cross_inversion} shows that recovering a non-C binary \emph{into C} beats recovering it into its own source language for four of the five languages (Fortran 29\% vs.\ 11\%, C++ 27\% vs.\ 12\%, Ada 13\% vs.\ 1\%, Rust 10\% vs.\ 1\%). Finally, all cases remain below the C$\to$C ceiling (59\% RE with iteration). To empirically validate the hypothesis that a performance drop occurs due to C-centric lifting by Ghidra, we evaluate the cross-language CIT-31b pipelines to ingest \texttt{asm}. Evidently, as seen in Fig. \ref{fig:rq3_model_iteration_role} and \ref{fig:same_cross_inversion}, pseudo-C adds 10.4\% RCI and 63.9\% RE for recovery of C source from C-derived assembly, but loses 8.4\% RCI and gains only 16.2\% RE for non-C languages. However, the improvement in cross-language RE for non-C to C over \texttt{asm} is still a motivation to use classical decompiler pseudo-C for source recovery. The full results of evaluations for \texttt{O0/O2} across 7 metrics derived from \texttt{gemma4-31b-it} and \texttt{qwen3.8-27b} are presented in Table~\ref{tab:rq4_language} in the Appendix.

\section{Key Findings and Open Problems}
\label{sec:key-findings-OP}
We consolidate our research findings based on our research questions and share open problems where applicable:
\begin{enumerate}
    \item \textbf{Value of the metrics (RQ1):} While Agent4Decompile leads the SOTA methods in terms of recompilability, functional similarity and byte-level semantics, FidelityGPT and SK2Decompile outperform in terms of readability and lexical similarity. This observation highlights the necessity of our multi-dimensional evaluation. 
    \item \textbf{Signature recovery failure (RQ1-4):} Comparison of RC vs RCI across our evaluations shows that while the SOTA (and recent LLMs in general) are highly capable at generating compilable code, they fail more often in generating the correct signature, which can be mitigated with harness-aware compiler feedback iterations.
    \item \textbf{Pseudo-C as a cross-architecture portability layer (RQ3):} Across the architectures, methods that ingest pseudo-C showcase higher fidelity compared to those that ingest assembly. Apparently, the uniformity of Ghidra's output absorbs the architectural shift that x86-64-assembly-trained models are sensitive to.
    \item \textbf{Performance drops for Thumb2 ISA (RQ2):} Even the best performing methods show worse performance on Thumb2 architectures such as armv7-m and armv8-m, compared to standard and non-Thumb2 embedded architectures. Since Thumb2 dominates deployed IoT firmware~\cite{tan2024sok} where source recovery is an important aspect of maintenance and upgrade of legacy firmware and vulnerability detection \cite{katpara2026firmware}, the fidelity of LLM-assisted methods in embedded systems is an open direction.
    \item \textbf{Drop with optimization (RQ2):} RE declines monotonically from \texttt{O0-O3}, and remains comparable for \texttt{Os} and \texttt{O2} since they share the same core functionality except code alignment and size-padding optimization \cite{gao2025decompilebench}. 
    \item \textbf{Off-the-shelf models and scale (RQ3):} Larger LLMs achieve higher overall fidelity, and methods that use general-purpose, off-the-shelf LLMs match the performance of those fine-tuned on pseudo-C. As a future work, clear benefits of fine-tuned models may arise for non-C. 
    \item \textbf{Control flow and multi-role benefits (RQ3):} Iteration improves RCI, BW, and RE over single-pass while sacrificing readability due to the scaffolding introduced to fix compilation errors. Multiple roles improve readability and lexical similarity, with marginal differences in compilability, semantic, and functional similarity.
    \item \textbf{Predominance of C (RQ4):} The fidelity of source recovery is worse on non-C-derived binaries. Cross-recovery to C, however, is better than recovering the original source. Ghidra pseudo-C offers limited gain over assembly compared to same-language recovery to C. These observations point towards \textit{cross-language decompilers, training and augmentation} as open directions. 
\end{enumerate}

\section{Threats to Validity}
ExeBench is a public benchmark and predates the training cutoffs of every model used in our work, and may be present in their pre-training dataset \cite{evertz-26-chasing}. However, our focus is on the comparative performance between methods. We also note that recovery quality is sensitive to prompt wording, input formatting, and decoding settings \cite{evertz-26-chasing}. SOTA methods are compared using their specified prompts, while ablations fix one prompt per task. We did not search over paraphrases, and different phrasings would likely move absolute scores. That being said, we publish all prompts along with source code. Lastly, we assume the availability and derivability of the function call interface due to the general availability of repository-level context in codebases. However, compiler interactions under the assumption of no repository-level context will show limited improvement in RCI.

\section{Conclusion}
This work presented the first SoK dedicated to LLM-assisted source code recovery. We created a novel design taxonomy and organized the published methods along five axes. We also consolidated the fragmented metric landscape in this topic into six key dimensions for a holistic evaluation. We evaluated representative SOTA methods on 45,000 samples built from 500 functions, 9~standard and embedded architectures, 5~optimization levels, and 2~symbol variants. Further, we ran a controlled, taxonomy-driven ablation of input representation, contextual enrichment, model scale, iterative feedback, and review roles, deriving key insights on impactful design choices. We also measured same- and cross-language (to C) recovery across mature and legacy source languages. Open problems for future work include the general fidelity, signature failures, and coverage of non-C and Thumb2-ISA. 

\section*{Ethical Considerations}
We make the following considerations: (1) While source recovery aids malware analysis, firmware vulnerability assessment, and legacy maintenance, it may also be used for IP theft and license circumvention. Our work measures and bounds the limitations of source recovery, and proposes no new method for the same. However, our evaluation harness may be used by people with malicious intent, and therefore, we urge readers to use our code ethically. (2) All our evaluations were conducted using permissively licensed C (Exebench). (3) All SOTA method evaluations were conducted using their original publicly sourced code from their respective authors. All deviations have been documented in the Appendix and our results were conducted on a unified evaluation set to ensure fair comparison.

\section*{LLM Usage Considerations}
LLMs were used for editorial purposes in this manuscript, and all outputs were inspected by the authors to ensure accuracy and originality. We used Anthropic's Claude Opus-5 and Fable-5.1, to assist with writing the code. The cross-language translations were done using Claude Fable-5.1, validated against the I/O test of the original C function for functional symmetry, and spot checked by the authors. All numerical results and figures were produced by our own pipeline, generated directly from the raw experimental outputs and checked against the underlying data. The authors have reviewed all AI-assisted code, and take full responsibility for the accuracy, originality, and integrity of this work. Further, LLMs play an integral role in our experiments since they are the focus of our evaluations and thus unavoidable to use. However, we carefully considered the extent of experiments required to support our claims: the SOTA evaluations study used a larger evaluation set for fair comparison across published methods; downstream ablations (RQ3/4) used only a subset (50 functions per language on one architecture and two optimizations) to derive their findings. Thus, our LLM usage has been under the overarching consideration of accountability, correctness, transparency, and responsibility. 

\section*{Open Science}

We release the complete artifact behind this work: the evaluation harness, study configurations, dataset, LLM generations, scored per-cell records, and the code that generates the tables and figures used in this paper. Details are available via the README.md file in the provided repository: https://anonymous.4open.science/r/decompbench-65CB/

\subsubsection{Code}
We provide the complete evaluation harness, including cross-compilation across nine architectures and five optimization levels, disassembly, Ghidra lifting, integrations for the eight evaluated recovery methods, and implementations of the seven evaluation metrics. Upstream methods are used verbatim with provenance, licenses, and with minimal changes to fit our study.
The study runners (scripts/inputrepr/, scripts/langstudy/, scripts/control\_flow/), the corpus sampler (halstead\_complexity/), the Docker stack (docker\_setup/), and the test suite are also provided. We provide one entry-point script per research question to run the models and reproduce the corresponding results: run\_rq2.sh, run\_rq3.sh, run\_rq4.sh. 
GPU resources are required to run the scripts as it serves the models used in the study. The inputs for RQ1 are obtained from the results of RQ2.

\subsubsection{Dataset}
The source of the sampled 500 functions dataset is Exebench's test\_real split. The functions are available in the repository for direct use with the harness. We provide a mapping to locate the original functions from the original split: sampled\_500\_sound\_by\_difficulty.json.

For RQ1 and RQ2, we use the 500 functions sampled dataset. The same functions are used for both research questions. For each method, we compile these functions into 45,000 binary samples using fixed toolchain versions to ensure reproducibility. For RQ3 and RQ4, we use a shared set of 50 functions derived from the samples 500, together with a 300-function multilingual dataset created by translating each of the 50 C functions into C++, Go, Rust, Fortran, and Ada using Claude Fable-5.1. The language translation process follows the specifications defined in scripts/langstudy/contracts/. Re-execution is evaluated using the input/output test suites generated using Exebench's testcase generation tool, located at decompbench/eval/generated\_io.

\subsubsection{Configs}
The configurations used for each research question are provided in the repository: arch\_safe\_exebench\_test\_real\_500.yaml (RQ1/RQ2), rq-input-representations.yaml (RQ3), rq-language-study.yaml (RQ4), rq-control-flow.yaml (the iteration and review arms), and study\_functions\_50.yaml, which RQ3 and RQ4 both use.

\subsubsection{Results}
The model generation results are stored under the \textbf{results} directory for RQ1 and RQ2, and \textbf{rq3\_ladder}, \textbf{rq4\_language}, \textbf{control\_flow} for RQ3/RQ4. We provide CSVs from our experiments to regenerate the tables and figures. To regenerate the results from the beginning, use the run\_rq scripts.

\subsubsection{Figures and tables}
\textbf{scripts/figures/make\_all\_figures.sh} rebuilds 16 figures and 6 tables from the released CSVs and the output is deterministic. The tables are written under the filenames the manuscript inputs.

\bibliographystyle{plainurl}
\bibliography{references}

\appendix

\begin{figure}[t]
    \centering
    \includegraphics[width=0.8\linewidth]{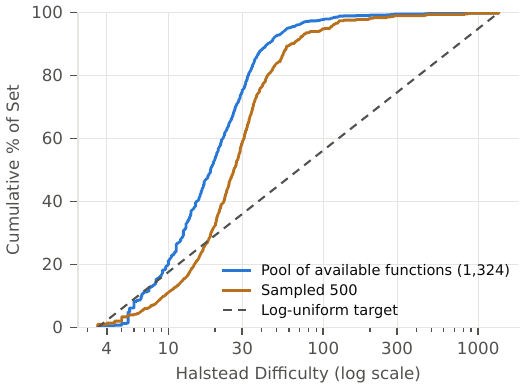}
    \caption{Halstead complexity distribution in the original Exebench and sampled split.}
    \label{fig:halstead}
\end{figure}

\begin{figure}[t]
    \centering
    \includegraphics[width=\linewidth]{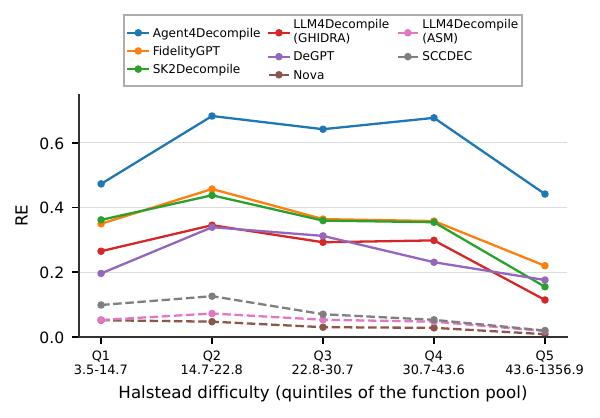}
    \caption{Method-wise performance against halstead groups.}
    \label{fig:halstead_performance}
\end{figure}

\begin{figure}[t]
    \centering
    \includegraphics[width=\linewidth]{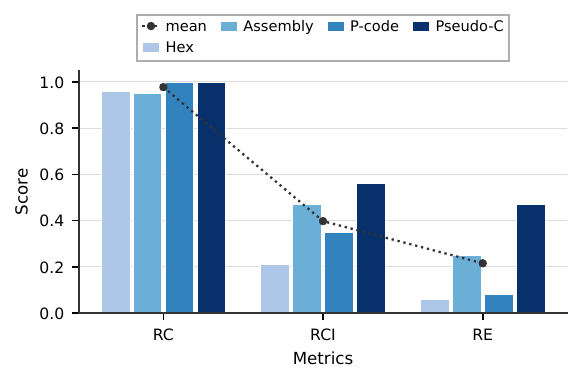}
    \caption{RC, RCI, and RE for base representation ablations.}
    \label{fig:representation_funnel}
\end{figure}

\begin{figure}[t]
    \centering
    \includegraphics[width=\linewidth]{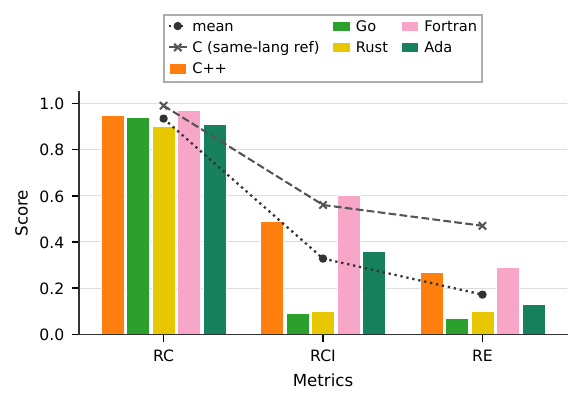}
    \caption{RC, RCI, and RE for cross-CIQ-31b across languages.}
    \label{fig:cross_CIQ_funnel}
\end{figure}

This appendix comprises the supplementary figures and tables for our readers' reference. Fig. \ref{fig:halstead} and \ref{fig:halstead_performance} depict the distribution of Halstead difficulty in the dataset and method-wise performance across difficulty bins, respectively. Fig. \ref{fig:representation_funnel} and \ref{fig:cross_CIQ_funnel} depict the nested relationship between RC, RCI, and RE across base input representation and cross-language recovery ablations. Tables \ref{tab:rq2_arch}, \ref{tab:rq2_opt}, and \ref{tab:rq2_symbols} contain the exhaustive results for SOTA methods across architectures, optimizations, and symbol variants. Tables \ref{tab:rq3_ladder_enrichment} and \ref{tab:rq3_scale_iteration} present the fine-grained results for the taxonomy-driven pipeline ablations. Lastly, Table \ref{tab:rq4_language} presents the results for same- and cross-language recovery.  

\input{appendix_tables/rq2_by_arch}

\input{appendix_tables/rq2_by_opt}

\input{appendix_tables/rq2_by_tier}

\begin{table*}[t]
\centering
\scriptsize
\caption{Input-representation ladder and enrichment channels, showing all seven metrics per optimization level.}
\label{tab:rq3_ladder_enrichment}
\begin{tabular}{ll*{14}{c}}
\toprule
Arm & Pipeline & \multicolumn{7}{c}{O0} & \multicolumn{7}{c}{O2} \\
\cmidrule(lr){3-9}\cmidrule(lr){10-16}
 &  & RCI & $\Delta_{\mathrm{R2I}}$ & ES & AST & DF & BW & RE & RCI & $\Delta_{\mathrm{R2I}}$ & ES & AST & DF & BW & RE \\
\midrule
\multirow{3}{*}{Hex} & SP-12b & 0.12 & -0.01 & 0.14 & 0.21 & 0.39 & 0.00 & 0.00 & 0.08 & 0.43 & 0.15 & 0.15 & 0.33 & 0.00 & 0.00 \\
 & SP-31b & 0.32 & -0.02 & 0.27 & 0.44 & 0.52 & 0.02 & 0.06 & 0.10 & 0.16 & 0.20 & 0.28 & 0.45 & 0.06 & 0.06 \\
 & SP-27b & 0.24 & 0.08 & 0.21 & 0.25 & 0.48 & 0.02 & 0.04 & 0.32 & 0.28 & 0.20 & 0.21 & 0.40 & 0.04 & 0.04 \\
\midrule
\multirow{3}{*}{Assembly} & SP-12b & 0.42 & 0.05 & 0.30 & 0.38 & 0.52 & 0.06 & 0.10 & 0.24 & 0.15 & 0.23 & 0.27 & 0.52 & 0.08 & 0.08 \\
 & SP-31b & 0.60 & 0.22 & 0.42 & 0.49 & 0.58 & 0.22 & 0.34 & 0.34 & 0.23 & 0.28 & 0.36 & 0.52 & 0.14 & 0.16 \\
 & SP-27b & 0.56 & 0.29 & 0.42 & 0.45 & 0.55 & 0.18 & 0.34 & 0.34 & 0.28 & 0.33 & 0.35 & 0.50 & 0.10 & 0.08 \\
\midrule
\multirow{3}{*}{P-code} & SP-12b & 0.24 & -0.01 & 0.25 & 0.32 & 0.50 & 0.00 & 0.04 & 0.26 & 0.03 & 0.23 & 0.28 & 0.49 & 0.00 & 0.02 \\
 & SP-31b & 0.38 & 0.22 & 0.31 & 0.39 & 0.53 & 0.02 & 0.08 & 0.32 & 0.12 & 0.27 & 0.36 & 0.53 & 0.02 & 0.08 \\
 & SP-27b & 0.28 & 0.35 & 0.33 & 0.35 & 0.51 & 0.02 & 0.06 & 0.36 & 0.31 & 0.31 & 0.34 & 0.52 & 0.04 & 0.04 \\
\midrule
\multirow{3}{*}{Pseudo-C} & SP-12b & 0.50 & 0.47 & 0.44 & 0.51 & 0.60 & 0.08 & 0.44 & 0.48 & 0.35 & 0.37 & 0.45 & 0.57 & 0.20 & 0.30 \\
 & SP-31b & 0.58 & 0.47 & 0.45 & 0.49 & 0.61 & 0.14 & 0.50 & 0.54 & 0.32 & 0.37 & 0.45 & 0.57 & 0.32 & 0.44 \\
 & SP-27b & 0.62 & 0.46 & 0.48 & 0.49 & 0.59 & 0.04 & 0.44 & 0.58 & 0.46 & 0.39 & 0.48 & 0.59 & 0.28 & 0.42 \\
\midrule
\multirow{2}{*}{+CFG} & SP-31b & 0.56 & 0.41 & 0.44 & 0.51 & 0.60 & 0.12 & 0.46 & 0.48 & 0.28 & 0.37 & 0.47 & 0.57 & 0.30 & 0.40 \\
 & SP-27b & 0.64 & 0.50 & 0.46 & 0.48 & 0.59 & 0.04 & 0.50 & 0.58 & 0.43 & 0.38 & 0.45 & 0.56 & 0.16 & 0.34 \\
\midrule
\multirow{2}{*}{+Frame/ABI} & SP-31b & 0.60 & 0.47 & 0.44 & 0.49 & 0.61 & 0.14 & 0.52 & 0.54 & 0.37 & 0.35 & 0.44 & 0.59 & 0.32 & 0.44 \\
 & SP-27b & 0.62 & 0.48 & 0.49 & 0.50 & 0.61 & 0.04 & 0.48 & 0.62 & 0.44 & 0.42 & 0.48 & 0.61 & 0.22 & 0.38 \\
\midrule
\multirow{2}{*}{+ROData} & SP-31b & 0.30 & 0.36 & 0.41 & 0.51 & 0.35 & 0.00 & 0.00 & 0.40 & 0.28 & 0.37 & 0.56 & 0.43 & 0.00 & 0.00 \\
 & SP-27b & 0.30 & 0.41 & 0.41 & 0.52 & 0.44 & 0.00 & 0.00 & 0.30 & 0.24 & 0.35 & 0.50 & 0.44 & 0.00 & 0.00 \\
\midrule
\multirow{2}{*}{+Callees} & SP-31b & 0.83 & 0.41 & 0.32 & 0.40 & 0.53 & 0.00 & 0.50 & 0.67 & 0.28 & 0.28 & 0.39 & 0.44 & 0.00 & 0.50 \\
 & SP-27b & 0.83 & 0.64 & 0.45 & 0.49 & 0.55 & 0.00 & 0.50 & 0.67 & 0.41 & 0.34 & 0.40 & 0.43 & 0.00 & 0.17 \\
\midrule
\multirow{2}{*}{+Data/BSS} & SP-31b & 0.00 & 0.65 & 0.23 & 0.43 & 0.45 & 0.00 & 0.00 & 0.00 & 0.36 & 0.24 & 0.19 & 0.18 & 0.00 & 0.00 \\
 & SP-27b & 1.00 & 0.09 & 0.36 & 0.59 & 0.82 & 0.00 & 1.00 & 0.00 & -0.60 & 0.26 & 0.22 & 0.18 & 0.00 & 0.00 \\
\midrule
\multirow{2}{*}{+All} & SP-31b & 0.64 & 0.40 & 0.44 & 0.48 & 0.60 & 0.12 & 0.54 & 0.62 & 0.31 & 0.36 & 0.45 & 0.58 & 0.32 & 0.48 \\
 & SP-27b & 0.66 & 0.48 & 0.48 & 0.49 & 0.62 & 0.06 & 0.46 & 0.68 & 0.41 & 0.41 & 0.47 & 0.59 & 0.18 & 0.38 \\
\bottomrule
\end{tabular}
\end{table*}

\begin{table*}[t]
\centering
\scriptsize
\caption{Impact of model scale, compiler iteration, and review role on the pseudo-C arm.}
\label{tab:rq3_scale_iteration}
\begin{tabular}{l*{14}{c}}
\toprule
Pipeline & \multicolumn{7}{c}{O0} & \multicolumn{7}{c}{O2} \\
\cmidrule(lr){2-8}\cmidrule(lr){9-15}
 & RCI & $\Delta_{\mathrm{R2I}}$ & ES & AST & DF & BW & RE & RCI & $\Delta_{\mathrm{R2I}}$ & ES & AST & DF & BW & RE \\
\midrule
SP-12b & 0.50 & 0.47 & 0.44 & 0.51 & 0.60 & 0.08 & 0.44 & 0.48 & 0.35 & 0.37 & 0.45 & 0.57 & 0.20 & 0.30 \\
SP-31b & 0.58 & 0.47 & 0.45 & 0.49 & 0.61 & 0.14 & 0.50 & 0.54 & 0.32 & 0.37 & 0.45 & 0.57 & 0.32 & 0.44 \\
CIT-31b & 0.76 & 0.39 & 0.43 & 0.52 & 0.60 & 0.18 & 0.62 & 0.72 & 0.26 & 0.36 & 0.47 & 0.57 & 0.32 & 0.56 \\
RR-31b & 0.56 & 0.52 & 0.43 & 0.47 & 0.60 & 0.12 & 0.54 & 0.52 & 0.37 & 0.35 & 0.43 & 0.57 & 0.32 & 0.48 \\
\bottomrule
\end{tabular}
\end{table*}

\begin{table*}[t]
\centering
\scriptsize
\caption{Same- and cross-language recovery across input representations and iteration, per source language. `--' marks structurally unmeasurable cells.}
\label{tab:rq4_language}
\begin{tabular}{ll*{14}{c}}
\toprule
Lang. & Arm & \multicolumn{7}{c}{O0} & \multicolumn{7}{c}{O2} \\
\cmidrule(lr){3-9}\cmidrule(lr){10-16}
 &  & RCI & $\Delta_{\mathrm{R2I}}$ & ES & AST & DF & BW & RE & RCI & $\Delta_{\mathrm{R2I}}$ & ES & AST & DF & BW & RE \\
\midrule
\multirow{4}{*}{C} & Same-SP-31b & 0.58 & 0.47 & 0.44 & 0.49 & 0.60 & 0.14 & 0.52 & 0.54 & 0.31 & 0.37 & 0.46 & 0.57 & 0.28 & 0.42 \\
 & Same-SP-27b & 0.60 & 0.52 & 0.48 & 0.50 & 0.60 & 0.06 & 0.46 & 0.62 & 0.46 & 0.39 & 0.46 & 0.59 & 0.24 & 0.38 \\
 & Same-asm-SP-31b & 0.54 & 0.17 & 0.41 & 0.46 & 0.59 & 0.14 & 0.30 & 0.34 & 0.19 & 0.27 & 0.35 & 0.54 & 0.16 & 0.16 \\
 & Same-asm-CIT-31b & 0.68 & 0.16 & 0.42 & 0.48 & 0.58 & 0.14 & 0.38 & 0.66 & 0.14 & 0.31 & 0.41 & 0.56 & 0.18 & 0.34 \\
\midrule
\multirow{10}{*}{C++} & Same-SP-31b & 0.98 & -- & 0.30 & 0.27 & 0.43 & -- & 0.16 & 0.90 & -- & 0.25 & 0.26 & 0.43 & -- & 0.06 \\
 & Same-SP-27b & 0.88 & -- & 0.33 & 0.28 & 0.43 & -- & 0.10 & 0.80 & -- & 0.30 & 0.30 & 0.43 & -- & 0.16 \\
 & Same-CIT-31b & 1.00 & -- & 0.30 & 0.27 & 0.43 & -- & 0.16 & 1.00 & -- & 0.26 & 0.26 & 0.42 & -- & 0.08 \\
 & Cross-SP-31b & 0.24 & 0.51 & 0.33 & 0.38 & 0.51 & 0.02 & 0.18 & 0.24 & 0.41 & 0.29 & 0.35 & 0.50 & 0.08 & 0.10 \\
 & Cross-SP-27b & 0.22 & 0.58 & 0.33 & 0.35 & 0.49 & 0.00 & 0.12 & 0.38 & 0.49 & 0.35 & 0.42 & 0.51 & 0.10 & 0.12 \\
 & Cross-CIT-31b & 0.40 & 0.42 & 0.31 & 0.38 & 0.50 & 0.04 & 0.28 & 0.58 & 0.25 & 0.28 & 0.41 & 0.44 & 0.12 & 0.26 \\
 & Same-asm-SP-31b & 0.86 & -- & 0.39 & 0.42 & 0.54 & -- & 0.14 & 0.90 & -- & 0.21 & 0.20 & 0.41 & -- & 0.06 \\
 & Same-asm-CIT-31b & 0.96 & -- & 0.39 & 0.42 & 0.54 & -- & 0.14 & 0.98 & -- & 0.21 & 0.20 & 0.41 & -- & 0.06 \\
 & Cross-asm-SP-31b & 0.28 & 0.24 & 0.32 & 0.37 & 0.51 & 0.04 & 0.22 & 0.16 & 0.32 & 0.25 & 0.30 & 0.48 & 0.04 & 0.04 \\
 & Cross-asm-CIT-31b & 0.52 & 0.24 & 0.33 & 0.39 & 0.50 & 0.04 & 0.30 & 0.44 & 0.21 & 0.26 & 0.33 & 0.48 & 0.06 & 0.16 \\
\midrule
\multirow{10}{*}{Rust} & Same-SP-31b & 0.58 & -- & 0.29 & 0.36 & 0.45 & -- & 0.02 & 0.78 & -- & 0.24 & 0.32 & 0.45 & -- & 0.00 \\
 & Same-SP-27b & 0.64 & -- & 0.22 & 0.29 & 0.45 & -- & 0.00 & 0.58 & -- & 0.20 & 0.27 & 0.43 & -- & 0.00 \\
 & Same-CIT-31b & 0.94 & -- & 0.29 & 0.35 & 0.46 & -- & 0.02 & 0.96 & -- & 0.24 & 0.32 & 0.45 & -- & 0.00 \\
 & Cross-SP-31b & 0.04 & 0.54 & 0.20 & 0.42 & 0.52 & 0.00 & 0.04 & 0.04 & 0.44 & 0.18 & 0.41 & 0.48 & 0.04 & 0.04 \\
 & Cross-SP-27b & 0.02 & 0.58 & 0.19 & 0.42 & 0.52 & 0.00 & 0.02 & 0.02 & 0.56 & 0.19 & 0.42 & 0.51 & 0.04 & 0.02 \\
 & Cross-CIT-31b & 0.10 & 0.30 & 0.17 & 0.44 & 0.48 & 0.02 & 0.10 & 0.10 & 0.31 & 0.17 & 0.43 & 0.46 & 0.04 & 0.10 \\
 & Same-asm-SP-31b & 0.70 & -- & 0.26 & 0.27 & 0.47 & -- & 0.00 & 0.84 & -- & 0.22 & 0.22 & 0.43 & -- & 0.00 \\
 & Same-asm-CIT-31b & 0.90 & -- & 0.25 & 0.26 & 0.47 & -- & 0.00 & 0.96 & -- & 0.22 & 0.22 & 0.44 & -- & 0.00 \\
 & Cross-asm-SP-31b & 0.04 & 0.43 & 0.19 & 0.37 & 0.57 & 0.00 & 0.02 & 0.04 & 0.46 & 0.18 & 0.34 & 0.52 & 0.04 & 0.02 \\
 & Cross-asm-CIT-31b & 0.12 & 0.34 & 0.19 & 0.40 & 0.56 & 0.02 & 0.08 & 0.12 & 0.38 & 0.20 & 0.39 & 0.53 & 0.04 & 0.08 \\
\midrule
\multirow{10}{*}{Go} & Same-SP-31b & 0.58 & -- & 0.32 & 0.48 & 0.40 & -- & 0.14 & 0.30 & -- & 0.26 & 0.44 & 0.41 & -- & 0.00 \\
 & Same-SP-27b & 0.10 & -- & 0.24 & 0.33 & 0.45 & -- & 0.00 & 0.12 & -- & 0.22 & 0.38 & 0.44 & -- & 0.00 \\
 & Same-CIT-31b & 0.94 & -- & 0.32 & 0.48 & 0.41 & -- & 0.22 & 0.96 & -- & 0.26 & 0.44 & 0.41 & -- & 0.10 \\
 & Cross-SP-31b & 0.10 & 0.53 & 0.24 & 0.42 & 0.49 & 0.00 & 0.10 & 0.04 & 0.60 & 0.22 & 0.37 & 0.45 & 0.00 & 0.04 \\
 & Cross-SP-27b & 0.08 & 0.52 & 0.25 & 0.44 & 0.47 & 0.00 & 0.00 & 0.04 & 0.49 & 0.21 & 0.36 & 0.43 & 0.04 & 0.04 \\
 & Cross-CIT-31b & 0.14 & 0.53 & 0.21 & 0.40 & 0.50 & 0.00 & 0.10 & 0.04 & 0.58 & 0.20 & 0.37 & 0.42 & 0.00 & 0.04 \\
 & Same-asm-SP-31b & 0.58 & -- & 0.27 & 0.37 & 0.49 & -- & 0.02 & 0.72 & -- & 0.22 & 0.33 & 0.44 & -- & 0.04 \\
 & Same-asm-CIT-31b & 0.92 & -- & 0.28 & 0.38 & 0.49 & -- & 0.04 & 0.96 & -- & 0.22 & 0.34 & 0.46 & -- & 0.04 \\
 & Cross-asm-SP-31b & 0.06 & 0.46 & 0.25 & 0.44 & 0.48 & 0.00 & 0.04 & 0.08 & 0.55 & 0.24 & 0.35 & 0.45 & 0.02 & 0.02 \\
 & Cross-asm-CIT-31b & 0.12 & 0.44 & 0.23 & 0.48 & 0.48 & 0.00 & 0.10 & 0.14 & 0.55 & 0.24 & 0.39 & 0.45 & 0.00 & 0.04 \\
\midrule
\multirow{10}{*}{Fortran} & Same-SP-31b & 0.90 & -- & 0.28 & -- & -- & -- & 0.14 & 0.80 & -- & 0.27 & -- & -- & -- & 0.04 \\
 & Same-SP-27b & 0.84 & -- & 0.32 & -- & -- & -- & 0.16 & 0.86 & -- & 0.30 & -- & -- & -- & 0.10 \\
 & Same-CIT-31b & 0.96 & -- & 0.29 & -- & -- & -- & 0.14 & 0.92 & -- & 0.27 & -- & -- & -- & 0.08 \\
 & Cross-SP-31b & 0.36 & 0.48 & 0.36 & 0.43 & 0.46 & 0.06 & 0.18 & 0.36 & 0.35 & 0.27 & 0.38 & 0.50 & 0.06 & 0.16 \\
 & Cross-SP-27b & 0.34 & 0.47 & 0.38 & 0.41 & 0.51 & 0.06 & 0.16 & 0.36 & 0.36 & 0.29 & 0.40 & 0.49 & 0.10 & 0.20 \\
 & Cross-CIT-31b & 0.62 & 0.33 & 0.36 & 0.46 & 0.48 & 0.08 & 0.30 & 0.58 & 0.27 & 0.28 & 0.38 & 0.47 & 0.08 & 0.28 \\
 & Same-asm-SP-31b & 0.82 & -- & 0.31 & -- & -- & -- & 0.20 & 0.62 & -- & 0.24 & -- & -- & -- & 0.16 \\
 & Same-asm-CIT-31b & 0.94 & -- & 0.31 & -- & -- & -- & 0.24 & 0.76 & -- & 0.25 & -- & -- & -- & 0.16 \\
 & Cross-asm-SP-31b & 0.50 & 0.20 & 0.35 & 0.35 & 0.51 & 0.10 & 0.22 & 0.34 & 0.25 & 0.24 & 0.33 & 0.48 & 0.10 & 0.14 \\
 & Cross-asm-CIT-31b & 0.64 & 0.18 & 0.37 & 0.41 & 0.51 & 0.12 & 0.28 & 0.64 & 0.18 & 0.26 & 0.40 & 0.50 & 0.10 & 0.20 \\
\midrule
\multirow{10}{*}{Ada} & Same-SP-31b & 0.16 & -- & 0.16 & -- & -- & -- & 0.00 & 0.16 & -- & 0.15 & -- & -- & -- & 0.02 \\
 & Same-SP-27b & 0.22 & -- & 0.13 & -- & -- & -- & 0.00 & 0.20 & -- & 0.13 & -- & -- & -- & 0.00 \\
 & Same-CIT-31b & 0.20 & -- & 0.17 & -- & -- & -- & 0.00 & 0.18 & -- & 0.15 & -- & -- & -- & 0.02 \\
 & Cross-SP-31b & 0.22 & 0.43 & 0.30 & 0.32 & 0.43 & 0.02 & 0.10 & 0.20 & 0.33 & 0.26 & 0.31 & 0.43 & 0.06 & 0.06 \\
 & Cross-SP-27b & 0.26 & 0.42 & 0.29 & 0.30 & 0.41 & 0.02 & 0.10 & 0.28 & 0.33 & 0.31 & 0.32 & 0.40 & 0.04 & 0.08 \\
 & Cross-CIT-31b & 0.36 & 0.38 & 0.27 & 0.34 & 0.43 & 0.04 & 0.14 & 0.36 & 0.22 & 0.27 & 0.35 & 0.45 & 0.10 & 0.12 \\
 & Same-asm-SP-31b & 0.08 & -- & 0.14 & -- & -- & -- & 0.00 & 0.16 & -- & 0.12 & -- & -- & -- & 0.00 \\
 & Same-asm-CIT-31b & 0.14 & -- & 0.15 & -- & -- & -- & 0.00 & 0.22 & -- & 0.13 & -- & -- & -- & 0.02 \\
 & Cross-asm-SP-31b & 0.30 & 0.14 & 0.26 & 0.27 & 0.48 & 0.02 & 0.08 & 0.16 & 0.17 & 0.22 & 0.24 & 0.46 & 0.04 & 0.04 \\
 & Cross-asm-CIT-31b & 0.42 & 0.14 & 0.27 & 0.28 & 0.49 & 0.02 & 0.12 & 0.42 & 0.11 & 0.23 & 0.28 & 0.45 & 0.06 & 0.12 \\
\bottomrule
\end{tabular}
\end{table*}

\end{document}

%% file: appendix_tables/rq2_by_arch.tex
\begin{sidewaystable*}[p]
\centering
\footnotesize
\setlength{\tabcolsep}{2pt}
\renewcommand{\arraystretch}{1.10}
\caption{All seven metrics per architecture, averaged over optimization levels and symbol tiers. }
\label{tab:rq2_arch}
\begin{tabular}{@{}ll*{28}{c}@{}}
\toprule
\multicolumn{30}{c}{\textbf{Standard targets}} \\
\midrule
Input & Pipeline & \multicolumn{7}{c}{x86-64} & \multicolumn{7}{c}{x86-32} & \multicolumn{7}{c}{AArch64} & \multicolumn{7}{c}{RISC-V64} \\
\cmidrule(lr){3-9}\cmidrule(lr){10-16}\cmidrule(lr){17-23}\cmidrule(lr){24-30}
 &  & RCI & $\Delta_{\mathrm{R2I}}$ & ES & AST & DF & BW & RE & RCI & $\Delta_{\mathrm{R2I}}$ & ES & AST & DF & BW & RE & RCI & $\Delta_{\mathrm{R2I}}$ & ES & AST & DF & BW & RE & RCI & $\Delta_{\mathrm{R2I}}$ & ES & AST & DF & BW & RE \\
\midrule
\multirow{3}{*}{Assembly} & LLM4Decompile (ASM) & 0.41 & -0.13 & 0.41 & 0.38 & 0.54 & 0.15 & 0.22 & 0.01 & -0.54 & 0.23 & 0.21 & 0.39 & 0.01 & 0.00 & 0.11 & -0.62 & 0.23 & 0.23 & 0.43 & 0.02 & 0.04 & 0.17 & -0.49 & 0.22 & 0.23 & 0.39 & 0.04 & 0.05 \\
 & Nova & 0.46 & 0.16 & 0.41 & 0.39 & 0.51 & 0.14 & 0.22 & 0.02 & -0.07 & 0.24 & 0.18 & 0.38 & 0.01 & 0.01 & 0.06 & -0.35 & 0.16 & 0.14 & 0.38 & 0.01 & 0.02 & 0.08 & -0.39 & 0.15 & 0.13 & 0.34 & 0.02 & 0.02 \\
 & SCCDEC & 0.46 & 0.11 & 0.43 & 0.42 & 0.54 & 0.22 & 0.26 & 0.03 & -0.35 & 0.24 & 0.23 & 0.38 & 0.02 & 0.02 & 0.16 & -0.36 & 0.25 & 0.25 & 0.40 & 0.04 & 0.06 & 0.21 & -0.41 & 0.21 & 0.24 & 0.36 & 0.06 & 0.07 \\
\midrule
\multirow{5}{*}{Pseudo-C} & LLM4Decompile (GHIDRA) & 0.51 & 0.13 & 0.43 & 0.42 & 0.54 & 0.20 & 0.32 & 0.52 & 0.14 & 0.42 & 0.41 & 0.55 & 0.17 & 0.33 & 0.43 & 0.06 & 0.41 & 0.41 & 0.54 & 0.17 & 0.29 & 0.21 & 0.16 & 0.42 & 0.42 & 0.48 & 0.06 & 0.13 \\
 & SK2Decompile & 0.62 & 0.33 & 0.45 & 0.46 & 0.55 & 0.24 & 0.39 & 0.64 & 0.32 & 0.45 & 0.47 & 0.54 & 0.23 & 0.42 & 0.53 & 0.33 & 0.44 & 0.46 & 0.54 & 0.21 & 0.36 & 0.30 & 0.35 & 0.44 & 0.46 & 0.48 & 0.11 & 0.19 \\
 & DeGPT & 0.38 & 0.29 & 0.36 & 0.38 & 0.53 & 0.13 & 0.31 & 0.41 & 0.30 & 0.36 & 0.38 & 0.53 & 0.11 & 0.34 & 0.32 & 0.29 & 0.34 & 0.37 & 0.53 & 0.12 & 0.28 & 0.22 & 0.32 & 0.36 & 0.38 & 0.48 & 0.09 & 0.19 \\
 & FidelityGPT & 0.53 & 0.40 & 0.41 & 0.42 & 0.55 & 0.20 & 0.39 & 0.57 & 0.40 & 0.41 & 0.42 & 0.56 & 0.19 & 0.43 & 0.49 & 0.42 & 0.40 & 0.43 & 0.56 & 0.20 & 0.36 & 0.41 & 0.42 & 0.39 & 0.41 & 0.50 & 0.14 & 0.32 \\
 & Agent4Decompile & 0.72 & 0.07 & 0.26 & 0.38 & 0.53 & 0.25 & 0.64 & 0.75 & 0.05 & 0.26 & 0.39 & 0.53 & 0.22 & 0.68 & 0.69 & 0.06 & 0.26 & 0.38 & 0.52 & 0.21 & 0.63 & 0.68 & 0.10 & 0.26 & 0.38 & 0.49 & 0.23 & 0.61 \\
\bottomrule
\end{tabular}
\par\vspace{1em}
\begin{tabular}{@{}ll*{21}{c}@{}}
\toprule
\multicolumn{23}{c}{\textbf{Embedded targets (1/2)}} \\
\midrule
Input & Pipeline & \multicolumn{7}{c}{ARMv7-A} & \multicolumn{7}{c}{ARMv8-A32} & \multicolumn{7}{c}{ARMv7-M} \\
\cmidrule(lr){3-9}\cmidrule(lr){10-16}\cmidrule(lr){17-23}
 &  & RCI & $\Delta_{\mathrm{R2I}}$ & ES & AST & DF & BW & RE & RCI & $\Delta_{\mathrm{R2I}}$ & ES & AST & DF & BW & RE & RCI & $\Delta_{\mathrm{R2I}}$ & ES & AST & DF & BW & RE \\
\midrule
\multirow{3}{*}{Assembly} & LLM4Decompile (ASM) & 0.13 & -0.45 & 0.21 & 0.22 & 0.40 & 0.03 & 0.03 & 0.14 & -0.46 & 0.21 & 0.22 & 0.40 & 0.03 & 0.03 & 0.07 & -0.51 & 0.18 & 0.18 & 0.40 & 0.01 & 0.02 \\
 & Nova & 0.02 & -0.35 & 0.14 & 0.11 & 0.35 & 0.01 & 0.00 & 0.02 & -0.34 & 0.14 & 0.12 & 0.35 & 0.01 & 0.00 & 0.02 & -0.43 & 0.11 & 0.09 & 0.35 & 0.01 & 0.00 \\
 & SCCDEC & 0.18 & -0.38 & 0.21 & 0.23 & 0.37 & 0.05 & 0.06 & 0.18 & -0.38 & 0.22 & 0.23 & 0.37 & 0.04 & 0.06 & 0.12 & -0.42 & 0.18 & 0.19 & 0.35 & 0.03 & 0.04 \\
\midrule
\multirow{5}{*}{Pseudo-C} & LLM4Decompile (GHIDRA) & 0.47 & 0.11 & 0.41 & 0.40 & 0.54 & 0.16 & 0.29 & 0.47 & 0.11 & 0.41 & 0.40 & 0.53 & 0.16 & 0.28 & 0.34 & 0.03 & 0.34 & 0.34 & 0.51 & 0.10 & 0.21 \\
 & SK2Decompile & 0.57 & 0.33 & 0.42 & 0.45 & 0.53 & 0.21 & 0.35 & 0.59 & 0.33 & 0.43 & 0.45 & 0.53 & 0.21 & 0.35 & 0.48 & 0.34 & 0.38 & 0.41 & 0.52 & 0.17 & 0.29 \\
 & DeGPT & 0.31 & 0.32 & 0.35 & 0.37 & 0.53 & 0.10 & 0.27 & 0.33 & 0.33 & 0.35 & 0.37 & 0.53 & 0.09 & 0.28 & 0.15 & 0.34 & 0.30 & 0.32 & 0.54 & 0.06 & 0.13 \\
 & FidelityGPT & 0.49 & 0.42 & 0.40 & 0.42 & 0.55 & 0.18 & 0.36 & 0.50 & 0.42 & 0.40 & 0.42 & 0.55 & 0.18 & 0.37 & 0.34 & 0.43 & 0.36 & 0.38 & 0.55 & 0.14 & 0.25 \\
 & Agent4Decompile & 0.66 & 0.04 & 0.26 & 0.38 & 0.51 & 0.21 & 0.61 & 0.67 & 0.04 & 0.26 & 0.37 & 0.52 & 0.21 & 0.61 & 0.44 & 0.03 & 0.23 & 0.34 & 0.52 & 0.21 & 0.40 \\
\bottomrule
\end{tabular}
\par\vspace{1em}
\begin{tabular}{@{}ll*{14}{c}@{}}
\toprule
\multicolumn{16}{c}{\textbf{Embedded targets (2/2)}} \\
\midrule
Input & Pipeline & \multicolumn{7}{c}{ARMv8-M} & \multicolumn{7}{c}{MIPS32LE} \\
\cmidrule(lr){3-9}\cmidrule(lr){10-16}
 &  & RCI & $\Delta_{\mathrm{R2I}}$ & ES & AST & DF & BW & RE & RCI & $\Delta_{\mathrm{R2I}}$ & ES & AST & DF & BW & RE \\
\midrule
\multirow{3}{*}{Assembly} & LLM4Decompile (ASM) & 0.09 & -0.45 & 0.18 & 0.17 & 0.40 & 0.01 & 0.02 & 0.11 & -0.62 & 0.18 & 0.20 & 0.39 & 0.02 & 0.03 \\
 & Nova & 0.02 & -0.46 & 0.11 & 0.09 & 0.34 & 0.01 & 0.00 & 0.10 & -0.41 & 0.14 & 0.13 & 0.32 & 0.03 & 0.03 \\
 & SCCDEC & 0.14 & -0.36 & 0.19 & 0.19 & 0.35 & 0.03 & 0.05 & 0.17 & -0.44 & 0.19 & 0.21 & 0.35 & 0.03 & 0.05 \\
\midrule
\multirow{5}{*}{Pseudo-C} & LLM4Decompile (GHIDRA) & 0.35 & 0.04 & 0.34 & 0.34 & 0.52 & 0.10 & 0.22 & 0.46 & 0.09 & 0.40 & 0.40 & 0.53 & 0.16 & 0.30 \\
 & SK2Decompile & 0.47 & 0.33 & 0.37 & 0.41 & 0.52 & 0.15 & 0.28 & 0.56 & 0.32 & 0.42 & 0.45 & 0.53 & 0.22 & 0.37 \\
 & DeGPT & 0.15 & 0.34 & 0.30 & 0.32 & 0.54 & 0.05 & 0.13 & 0.39 & 0.29 & 0.35 & 0.38 & 0.53 & 0.13 & 0.33 \\
 & FidelityGPT & 0.34 & 0.43 & 0.36 & 0.38 & 0.55 & 0.14 & 0.25 & 0.53 & 0.41 & 0.40 & 0.41 & 0.55 & 0.21 & 0.41 \\
 & Agent4Decompile & 0.44 & 0.02 & 0.23 & 0.34 & 0.51 & 0.21 & 0.41 & 0.72 & 0.04 & 0.26 & 0.38 & 0.52 & 0.26 & 0.66 \\
\bottomrule
\end{tabular}
\end{sidewaystable*}

%% file: appendix_tables/rq2_by_opt.tex
\begin{sidewaystable*}[p]
\centering
\footnotesize
\setlength{\tabcolsep}{2pt}
\renewcommand{\arraystretch}{1.10}
\caption{All seven metrics per optimization level, averaged over nine architectures and both symbol tiers. }
\label{tab:rq2_opt}
\begin{tabular}{@{}ll*{21}{c}@{}}
\toprule
Input & Pipeline & \multicolumn{7}{c}{O0} & \multicolumn{7}{c}{O1} & \multicolumn{7}{c}{O2} \\
\cmidrule(lr){3-9}\cmidrule(lr){10-16}\cmidrule(lr){17-23}
 &  & RCI & $\Delta_{\mathrm{R2I}}$ & ES & AST & DF & BW & RE & RCI & $\Delta_{\mathrm{R2I}}$ & ES & AST & DF & BW & RE & RCI & $\Delta_{\mathrm{R2I}}$ & ES & AST & DF & BW & RE \\
\midrule
\multirow{3}{*}{Assembly} & LLM4Decompile (ASM) & 0.16 & -0.52 & 0.22 & 0.21 & 0.41 & 0.04 & 0.07 & 0.14 & -0.45 & 0.24 & 0.25 & 0.43 & 0.04 & 0.05 & 0.13 & -0.49 & 0.22 & 0.23 & 0.41 & 0.03 & 0.04 \\
 & Nova & 0.07 & -0.48 & 0.11 & 0.10 & 0.33 & 0.02 & 0.04 & 0.10 & -0.26 & 0.19 & 0.17 & 0.39 & 0.03 & 0.03 & 0.08 & -0.25 & 0.19 & 0.16 & 0.38 & 0.03 & 0.03 \\
 & SCCDEC & 0.20 & -0.44 & 0.22 & 0.23 & 0.37 & 0.05 & 0.09 & 0.19 & -0.30 & 0.25 & 0.26 & 0.40 & 0.07 & 0.08 & 0.17 & -0.33 & 0.23 & 0.24 & 0.38 & 0.06 & 0.07 \\
\midrule
\multirow{5}{*}{Pseudo-C} & LLM4Decompile (GHIDRA) & 0.47 & 0.17 & 0.44 & 0.43 & 0.56 & 0.11 & 0.32 & 0.42 & 0.11 & 0.41 & 0.40 & 0.53 & 0.17 & 0.26 & 0.40 & 0.07 & 0.39 & 0.38 & 0.52 & 0.14 & 0.24 \\
 & SK2Decompile & 0.57 & 0.38 & 0.49 & 0.50 & 0.58 & 0.13 & 0.42 & 0.54 & 0.33 & 0.43 & 0.45 & 0.53 & 0.23 & 0.33 & 0.52 & 0.32 & 0.41 & 0.43 & 0.51 & 0.20 & 0.30 \\
 & DeGPT & 0.33 & 0.32 & 0.37 & 0.37 & 0.55 & 0.05 & 0.30 & 0.31 & 0.32 & 0.34 & 0.37 & 0.52 & 0.11 & 0.26 & 0.29 & 0.33 & 0.33 & 0.36 & 0.52 & 0.10 & 0.23 \\
 & FidelityGPT & 0.51 & 0.47 & 0.45 & 0.44 & 0.60 & 0.11 & 0.42 & 0.48 & 0.41 & 0.39 & 0.41 & 0.55 & 0.20 & 0.36 & 0.45 & 0.40 & 0.37 & 0.40 & 0.53 & 0.18 & 0.32 \\
 & Agent4Decompile & 0.69 & 0.13 & 0.29 & 0.39 & 0.55 & 0.10 & 0.64 & 0.65 & 0.04 & 0.25 & 0.37 & 0.52 & 0.28 & 0.59 & 0.62 & 0.02 & 0.24 & 0.36 & 0.51 & 0.24 & 0.56 \\
\bottomrule
\end{tabular}
\par\vspace{1em}
\begin{tabular}{@{}ll*{14}{c}@{}}
\toprule
Input & Pipeline & \multicolumn{7}{c}{O3} & \multicolumn{7}{c}{Os} \\
\cmidrule(lr){3-9}\cmidrule(lr){10-16}
 &  & RCI & $\Delta_{\mathrm{R2I}}$ & ES & AST & DF & BW & RE & RCI & $\Delta_{\mathrm{R2I}}$ & ES & AST & DF & BW & RE \\
\midrule
\multirow{3}{*}{Assembly} & LLM4Decompile (ASM) & 0.13 & -0.50 & 0.21 & 0.22 & 0.41 & 0.03 & 0.04 & 0.14 & -0.42 & 0.24 & 0.23 & 0.42 & 0.04 & 0.04 \\
 & Nova & 0.09 & -0.23 & 0.19 & 0.16 & 0.37 & 0.02 & 0.03 & 0.10 & -0.24 & 0.20 & 0.18 & 0.37 & 0.03 & 0.03 \\
 & SCCDEC & 0.17 & -0.34 & 0.22 & 0.23 & 0.38 & 0.06 & 0.07 & 0.19 & -0.25 & 0.26 & 0.25 & 0.40 & 0.06 & 0.07 \\
\midrule
\multirow{5}{*}{Pseudo-C} & LLM4Decompile (GHIDRA) & 0.40 & 0.06 & 0.36 & 0.37 & 0.50 & 0.14 & 0.25 & 0.40 & 0.08 & 0.39 & 0.39 & 0.53 & 0.15 & 0.24 \\
 & SK2Decompile & 0.51 & 0.31 & 0.38 & 0.41 & 0.50 & 0.20 & 0.30 & 0.51 & 0.31 & 0.42 & 0.44 & 0.52 & 0.21 & 0.32 \\
 & DeGPT & 0.26 & 0.29 & 0.33 & 0.36 & 0.52 & 0.11 & 0.22 & 0.29 & 0.30 & 0.34 & 0.36 & 0.52 & 0.11 & 0.25 \\
 & FidelityGPT & 0.44 & 0.39 & 0.35 & 0.38 & 0.52 & 0.18 & 0.32 & 0.45 & 0.41 & 0.39 & 0.41 & 0.54 & 0.20 & 0.33 \\
 & Agent4Decompile & 0.60 & 0.02 & 0.24 & 0.36 & 0.51 & 0.25 & 0.54 & 0.64 & 0.04 & 0.25 & 0.37 & 0.51 & 0.25 & 0.58 \\
\bottomrule
\end{tabular}
\end{sidewaystable*}

%% file: appendix_tables/rq2_by_tier.tex
\begin{table*}[t]
\centering
\footnotesize
\renewcommand{\arraystretch}{1.10}
\setlength{\tabcolsep}{2.8pt}
\caption{All seven metrics per symbol tier, averaged over the nine architectures and the five optimization levels. }
\label{tab:rq2_symbols}
\begin{tabular}{ll*{14}{c}}
\toprule
Input & Pipeline & \multicolumn{7}{c}{Unstripped} & \multicolumn{7}{c}{Stripped} \\
\cmidrule(lr){3-9}\cmidrule(lr){10-16}
 &  & RCI & $\Delta_{\mathrm{R2I}}$ & ES & AST & DF & BW & RE & RCI & $\Delta_{\mathrm{R2I}}$ & ES & AST & DF & BW & RE \\
\midrule
\multirow{3}{*}{Assembly} & LLM4Decompile (ASM) & 0.18 & -0.38 & 0.24 & 0.24 & 0.42 & 0.04 & 0.06 & 0.10 & -0.57 & 0.21 & 0.21 & 0.41 & 0.03 & 0.04 \\
 & Nova & 0.09 & -0.26 & 0.18 & 0.16 & 0.37 & 0.03 & 0.03 & 0.09 & -0.32 & 0.17 & 0.15 & 0.36 & 0.03 & 0.03 \\
 & SCCDEC & 0.24 & -0.22 & 0.27 & 0.28 & 0.42 & 0.07 & 0.09 & 0.13 & -0.45 & 0.20 & 0.21 & 0.36 & 0.05 & 0.05 \\
\midrule
\multirow{5}{*}{Pseudo-C} & LLM4Decompile (GHIDRA) & 0.45 & 0.19 & 0.43 & 0.42 & 0.52 & 0.18 & 0.29 & 0.38 & 0.01 & 0.37 & 0.37 & 0.53 & 0.11 & 0.23 \\
 & SK2Decompile & 0.53 & 0.34 & 0.44 & 0.45 & 0.52 & 0.19 & 0.33 & 0.52 & 0.32 & 0.41 & 0.44 & 0.54 & 0.19 & 0.34 \\
 & DeGPT & 0.29 & 0.32 & 0.34 & 0.36 & 0.52 & 0.09 & 0.24 & 0.31 & 0.31 & 0.34 & 0.36 & 0.53 & 0.10 & 0.26 \\
 & FidelityGPT & 0.47 & 0.44 & 0.40 & 0.41 & 0.55 & 0.17 & 0.35 & 0.47 & 0.40 & 0.38 & 0.40 & 0.55 & 0.18 & 0.35 \\
 & Agent4Decompile & 0.65 & 0.05 & 0.25 & 0.37 & 0.51 & 0.22 & 0.59 & 0.63 & 0.05 & 0.25 & 0.37 & 0.52 & 0.22 & 0.58 \\
\bottomrule
\end{tabular}
\end{table*}